\documentclass[superscriptaddress,floatfix,pra,twocolumn,amsmath,amssymb]{revtex4-2}
\usepackage{caption}
\usepackage{subcaption}
\usepackage{float}
\usepackage{lipsum}
\usepackage{graphicx}
\usepackage{xcolor}
\usepackage{mathrsfs}
\usepackage{mathtools}
\usepackage{braket}
\usepackage{multirow}
\usepackage{algorithm}  
\usepackage{algpseudocode}
\usepackage{placeins}
\usepackage{enumitem}  
\usepackage{placeins}
\usepackage{afterpage}
\usepackage{url}
\usepackage{booktabs} 
\usepackage{bm}
\usepackage[margin=1in]{geometry}
\usepackage{adjustbox} 
\usepackage{hyperref} 
\hypersetup{allcolors=blue}

\newcommand{\tr}{\mathrm{tr}}
\DeclareMathOperator*{\argmin}{arg\,min}
\newcommand{\bracket}[1]{\langle 1\rangle}

\algrenewcommand\algorithmicrequire{\State\textbf{Input:}}
\algrenewcommand\algorithmicensure {\State\textbf{Output:}}

\begin{document}

\title{Quantum Engineering of Qudits with Interpretable Machine Learning}

\author{Yule Mayevsky}
\address{Quantum Photonics Laboratory and Centre for Quantum Computation and Communication Technology, RMIT University, Melbourne, VIC 3000, Australia}

\author{Akram Youssry}
\address{Quantum Photonics Laboratory and Centre for Quantum Computation and Communication Technology, RMIT University, Melbourne, VIC 3000, Australia}
\email{akram.youssry.mohamed@rmit.edu.au}

\author{Ritik Sareen}
\address{Quantum Photonics Laboratory and Centre for Quantum Computation and Communication Technology, RMIT University, Melbourne, VIC 3000, Australia}

\author{Gerardo A. Paz-Silva}
\address{Centre for Quantum Dynamics and  Centre for Quantum Computation and Communication Technology, Griffith University, Brisbane, Queensland 4111, Australia}

\author{Alberto Peruzzo}
\address{Quantum Photonics Laboratory and Centre for Quantum Computation and Communication Technology, RMIT University, Melbourne, VIC 3000, Australia}
\address{Quandela, Massy, France}

\begin{abstract}
    In this paper, we show innovation in methods regarding extending graybox to higher dimensions:
    \begin{enumerate}
        \item We introduce a new design of graybox that extends to and dimension $d$
        \item We show how to choose basis for qudits that are suitable for graybox
        \item We show graybox learning of noiseless and noisy qubit, 2qubit and qudits.
        \item We show fidelity to implement some SU(d) operations
        \item We show fidelity for 2-level subspace operation addressing the leakage problem
        \item propose unboxing approach,  uses a local Taylor expansion that obtains an analytical approximation of the noise operator yeilding a new control cost function for control optimisations.
    \end{enumerate}
\end{abstract}

\begin{abstract}
Higher-dimensional quantum systems (qudits) offer advantages in information encoding, error resilience, and compact gate implementations, and naturally arise in platforms such as superconducting and solid-state systems. However, realistic conditions such as non-Markovian noise, non-ideal pulses, and beyond rotating wave approximation (RWA) dynamics, pose significant challenges for controlling and characterizing qudits. In this work, we present a machine-learning-based graybox framework for the control and noise characterization of qudits with arbitrary dimension, extending recent methods developed for single-qubit systems. Additionally, we introduce a local analytic expansion that enables interpretable modelling of the noise dynamics, providing a structured and efficient way to simulate system behaviour and compare different noise models. This interpretability feature allows us to to understand the mechanisms underlying successful control strategies; and opens the way for developing methods for distinguishing noise sources with similar effects. We demonstrate high-fidelity implementations of both global unitary operations as well as two-level subspace gates. Our work establishes a foundation for scalable and interpretable quantum control techniques applicable to both NISQ devices and finite-dimensional quantum systems, enhancing the performance of next-generation quantum technologies.

\end{abstract}

\maketitle

High-fidelity quantum control is essential for realizing scalable quantum technologies, including quantum computation, communication, and metrology. While most quantum processors are designed around qubits—two-level systems—many physical implementations naturally exhibit multi-level dynamics. These qudits, with Hilbert space dimension $ d > 2 $, arise in a wide range of platforms, including photonic systems \cite{Erhard2020}, superconducting circuits \cite{Neeley2009, goss2022high}, trapped ions \cite{Blatt2012}, ion implanted \cite{Mourik2018}, and nitrogen-vacancy (NV)  \cite{doherty2013nitrogen}. Exploiting these higher-dimensional structures offers several potential advantages. Recent work have demonstrated controllable multi-level operations in solid-state systems \cite{Fuentes2024, zhou2017holonomic}, as well as the ability to encode and manipulate continuous-variable dynamics approximated within a finite-dimensional Hilbert spaces \cite{Yu2024}. These systems enable more compact and expressive quantum circuits while offering advantages for quantum communication and cryptographic protocols \cite{Erhard2020, Wang2020, Cerf2002, Bechmann2000}, and improved fault-tolerance through more efficient error correction schemes \cite{Campbell2014, gottesman1998heisenberg}. Qudits also provide a unique platform for exploring fundamental phenomena such as quantum chaos in single-particle systems \cite{Mourik2018}.

Despite these benefits, control of qudits remains significantly more challenging than control of qubits. As dimensionality increases, so does the complexity of the system's dynamics: higher-order couplings, leakage channels, off-resonant transitions, and spectral crowding become more prevalent. These effects are further exacerbated by non-ideal control pulses \cite{Motzoi2009,Schutjens2013,Warren1994} and non-Markovian noise \cite{morris2022quantifying, white2020demonstration, giarmatzi2023multi}, which are increasingly common in near-term quantum hardware. In such regimes, widely used approximations—such as the rotating wave approximation (RWA)—often break down \cite{yang2024highfidelityczgatesdouble,  burgarthtaming, burgarthquantifying}, rendering traditional analytical and numerical control approaches inadequate.

To address these challenges, hybrid machine-learning approaches to quantum control have attracted growing interest. In particular, the Graybox (graybox) framework combines a physics-informed whitebox model with a trainable blackbox component to learn control protocols directly from data \cite{Rabitz2000, Fosel2018}. Previous applications of Graybox to single-qubit systems have demonstrated high-fidelity gate synthesis under realistic noise \cite{youssry2023noise,youssry2022multi,auza2024quantum}, using a straightforward cost function: the mean squared error (MSE) between a target gate and the model's predicted gate evolution. However, these results were limited to $ d = 2 $ systems.

Moreover, optimizing only for an MSE‐based cost function offers limited insight into how or why these models succeed—motivating a closer examination of their transparency through intrinsic interpretability versus post-hoc interpretability (explainability). 
Some of the early attempts to provide such a framework in quantum machine learning include the work in \cite{pira, pira2025enhanced,ran2023tensor,power2024feature}. Recent work in classical machine learning has emphasized the distinction between both paradigms \cite{murdoch2019definitions}. Intrinsically interpretable models expose their internal structure and parameters in human-understandable terms—such as linear models, decision trees, or physics-informed architectures. By contrast, post-hoc interpretability (which we deem ``explainability'') applies to blackbox models like deep neural networks, where insight is recovered only after training through surrogate approximations or feature attribution. This distinction reflects a fundamental trade-off between transparency and expressiveness, particularly in high-dimensional, noise-prone domains like quantum control. In response, a growing literature has proposed hybrid frameworks that embed domain knowledge at key model interfaces while preserving the flexibility of blackbox learners \cite{ferry2023learning,wang2021hybrid,baier2021hybrid}.

In this work, we extend the graybox approach to qudits. This is achieved by a redesigned blackbox model capable of representing higher-dimensional dynamics, as well as an improved whitebox component that enables a new input encoding scheme that facilitates efficient dataset generation. To validate our design, we generated synthetic datasets for a qubit and a qutrit system subject to strong classical non-Markovian noise. We then applied our Graybox-based optimization of universal gate sets for both qutrits and qubits yielding high-fidelity solutions. We find that the MSE-based cost function continues to yield high-fidelity control in qudit systems, despite the increased dimensionality and noise complexity. 

Our second contribution is providing a hybrid interpretable-explainable framework by extracting structured insight into how control pulses influence the system dynamics. At the heart of this framework is a local analytic expansion that quantifies the system’s response to pulse perturbations and reveals how deviations from ideal behavior arise from environmental noise and control imperfections. Our proposed model constrains the output layer to predict physically meaningful observables, while allowing the internal representation to remain latent. 

Our method provides a universal qudit system identification and control framework that can deal simultaneously with non-RWA dynamics, non-ideal pulse shapes, and non-Markovian noise. The framework supports structured, interpretable predictions without sacrificing model capacity—enabling downstream analysis grounded in physics. Beyond improving control fidelity, our results lay the foundation for scalable, noise-aware quantum system identification and gate design, bridging high performance with physical transparency.

\section{Problem setting}
In this paper, we study quantum systems defined on a $d$-dimensional Hilbert space, with dynamics governed by the time-dependent Hamiltonian:

\begin{align}
H(t) = H_{\text{drift}} + H_{\text{control}}(t) + H_{\text{noise}}(t),
\end{align}

where $H_{\text{drift}}$ is a known static drift term, $H_{\text{control}}(\mathrm{t})$ is a known control Hamiltonian, and $H_{\text{noise}}(\mathrm{t})$ is an unknown noise term that may include classical stochastic processes and quantum bath interactions \cite{youssry2020characterization, youssry2022multi}. The control Hamiltonian consists of $N_C$ parameterized control fields $f_i(t; \vec{\theta}_i)$ with parameters $\vec{\theta_i}$ acting via system operators $X_i$:

\begin{align}
H_{\text{control}}(t) = \sum_{i=0}^{N_C} f_i(t; \vec{\theta}_i) X_i.
\end{align}

The concatenation of the parameters $\vec{\theta}_i$ of each control field $X_i$ is denoted by $\vec{\theta}$. The system evolves from an initial state $\rho$ under $H(t)$ until time $t = T$, at which point an observable $O$ is measured. This is repeated to create an informationally-complete set of observables $\{\rho_i, O_j\}_{i,j}$. The concatenation of the expectation values of this set is denoted by $\vec{E}$. The task is to optimize the control parameters $\vec{\theta}$ such that the resulting evolution implements a target unitary gate $G$.

Here, we express the dynamics of the quantum system using the noise operator formalism from \cite{Gerardo,youssry2020characterization}, which expresses the expectation value of $O$ as:

\begin{align}
    \braket{O(T)} = \tr{\left( V_O(T) U_0(T) \rho U_0^\dagger(T) O \right)},
    \label{equ:Vo}
\end{align}

where $U_0(T) = \mathcal{T}e^{-i\int_0^T H_0(s) ds}$ is the unitary evolution under the known Hamiltonian $H_0(t) = H_{\text{drift}} + H_{\text{control}}(t)$, and $V_O(T)$ captures the effect of noise and control. While $V_O(T)$ is inaccessible and analytically intractable in general, it provides a powerful tool for analyzing system-environment interactions and their influence on quantum gates. 

To make $V_O(T)$ learnable, we employ a parameterized representation constrained by physical principles as will be shown in the next section. This allows $V_O(T)$ to be inferred from data while preserving its physical constraints. 
As such, we adopt a model-based graybox approach, leveraging a data-driven machine learning (ML) framework. The model inputs are the control parameters $\vec{\theta}$ and the outputs are the predicted expectation values $\vec{E}$ of 
a set of observables. A dataset of $(\vec{\theta}, \vec{E})$ pairs is generated by measuring expectation values for randomized pulse sequences. The model is trained to minimize prediction error (e.g., MSE) and validated on testing data.

Once trained, the model is integrated into a feedback loop to optimize $\vec{\theta}$ for implementing the target $G$. At each iteration of the optimization, the model prediction of the measurement outcome $\hat{E}(\vec{\theta}; \rho, O)$ corresponding to initial state $\rho$ and measurement operator $O$, is compared to the ideal expectation under $G$, and the MSE over the whole set of observables is calculated and optimized over. In other words, the cost function optimization is:
\begin{align}
    C(\vec{\theta}; G) = \sum_{\rho, O} \left( \tr\left( G \rho G^\dagger O \right) - \hat{E}(\vec{\theta}; \rho, O) \right)^2.
    \label{equ:msecostfunction}
\end{align}
Formally, the optimal solution is given by
\begin{align}
    \vec{\theta}^*= \argmin_{\vec{\theta} \in \mathcal{R}} C(\vec{\theta};G), 
\end{align}
where $\mathcal{R}$ represents the bounds for the allowable range of the control parameters $\vec{\theta}$. The optimized pulse sequence is then evaluated using the process fidelity metric:
\begin{align}
\text{Fidelity} = \tr \left( \sqrt{\sqrt{J_{\text{target}}} J_{\text{actual}} \sqrt{J_{\text{target}}}} \right),
\end{align}
where $J_{\text{target}}$ and $J_{\text{actual}}$ are the Choi representations of the target and the optimized operations.

In the next section, we present key components in building the graybox model for qudits: the parameterization of $V_O$, basis selection, handling of non-invertible observables, and the generalization properties of the model. This enables both high-fidelity gate design and structured noise characterization in open-system quantum control.

\section{Methodology}
\subsection{Finite dimensional noise operator}
\subsubsection{Parameterization of the noise operators}
The proposal in \cite{youssry2020characterization} is limited to single-qubit systems. In this paper we propose a new parameterization that can generalize to arbitrary system of dimension $d$. The noise operator of an observable $O$ can be expressed as \cite{youssry2020characterization}
\begin{align} 
    V_O = O^{-1} W_O.
    \label{equ: Voparam1}
\end{align}
The operator $W_O$ has to satisfy the constraints, that is Hermitian, with bounded eigenvalues and bounded trace:
\begin{enumerate}
    \item $W_O = W_O^{\dagger}$.
    \item The eigenvalues $\lambda_i$ of the operator $W_O$ satisfy that $|\lambda_i|\le d$, where $d$ is the dimensionality of the Hilbert space.
    \item $|\tr{(W_O)}|\le d \implies \sum_{i=1}^{d} \lambda_i \le d$.
\end{enumerate}
To satisfy those constraints we can use the eigendecomposition of $W_O$ operator,
\begin{align}
    W_O = Q D Q^{\dagger}.
    \label{equ: Voparam2}
\end{align}
The matrix $Q$ is a general $d \times d$ unitary. This can be parametrized using $2\times 2$ subunitary decomposition \cite{Nchuang}. This decomposition has a lot of applications in optics \cite{reck,youssry2025universal}, and the idea behind it is that we can decompose the general interaction between $d$ modes as a sequence of all possible interactions between 2 modes. In particular, a unitary matrix $Q$ can be written as product of $d(d-1)/2$ unitaries,
\begin{align}
    Q = \prod_{p=1}^{d} \prod_{q=p+1}^{d} U_{pq},
    \label{equ:Q}
\end{align}
where each unitary $U_{pq}$ is constructed by taking the identity matrix $I_{d \times d}$, and replace the elements located at locations $(p,p)$, $(p,q)$, $(q,p)$, and $(q,q)$ with the elements $(1,1)$, $(1,2)$, $(2,1)$, $(2,2)$ of a $2\times 2$ unitary. A general $2 \times 2$ unitary can be expressed as
\begin{align}
    U_{2\times2} = \begin{pmatrix}
        r e^{i\Theta} & \sqrt{1-r^2}e^{i \Psi} \\
        -\sqrt{1-r^2}e^{-i\Psi} & r e^{-i\Theta}
    \end{pmatrix},
\end{align}
with $0\le r \le 1$, $0 \le \Theta \le 2\pi$, and $0 \le \Psi \le 2\pi$. These parameters are all bounded. Thus, the total number of independent parameters needed for $Q$ is $3d(d-1)/2$. Now, for the eignevalues which form the diagonal elements of $D$, assume we have a set of parameters $x_1, x_2, \cdots x_d$ that satisfy that $-1 \le x_i \le 1$, and another set of parameters $p_1, p_2, \cdots p_d$ that form a probability distribution, i.e. $0 \le p_i \le 1$, and $\sum_i p_i = 1$. Define
$z_i = x_i p_i$, then we have that
\begin{align}
    |z_i| = |x_i p_i| = |x_i| |p_i| \le 1 \\
\end{align}
Let $\min_i x_i := x_{\min}$, and $\max_i x_i := x_{\max}$, then
\begin{align}
    x_{\min} &\le x_i \le x_{\max} \\
      x_{\min} &\le \sum_i  p_i x_i \le x_{\max}
\end{align}
Since we chose $-1 \le x_i \le 1$, this proves that $-1 \le z_i \le 1$. Therefore, we can choose the eigenvalues to be $z_i$. In other words,
\begin{align}
    D = d\begin{pmatrix}
        p_1 x_1  & 0 & \cdots & 0 \\ 0 & p_2 x_2  & \cdots & 0 \\ \vdots & \vdots & \ddots & \vdots\\
        0 & 0 & \cdots & p_dx_d 
    \end{pmatrix},
    \label{equ: D}
\end{align}
This parameterization enables the graybox framework to approximate the noise operator $ V_O(T) $ using learnable parameters that can be generated by a neural network. Specifically, the bounded and structured nature of the parameters ($ r, \Theta, \Psi $, and  $ z_i $) ensures compatibility with neural network outputs and facilitates efficient training. By incorporating this parameterization, the graybox framework can model the complex dynamics of $ V_O(T) $ while adhering to the physical constraints of quantum systems.

The formulation $ V_O = O^{-1} W_O $ provides a useful parameterization for invertible observables. However, many observables in quantum systems are non-invertible, which disrupts this parameterization. We address this challenge next.

\subsubsection{Handling non-invertible observables}

A non-invertible observable $O$ will result in an undefined $V_O$ operator. However, this can be addressed by applying a linear transformation with parameters $a,b$ to the observable $O$ yielding, 
\begin{align}
\tilde{O} = a O + b I_{d\times d},
\end{align}
The parameter $ b $ shifts the eigenvalues of $ O $ such that $ \tilde{O} $ becomes invertible \cite{grego}. Specifically, $ b $ is chosen to lie within the range defined by the minimum and maximum eigenvalues of $ O $ after scaling by $ a $. This ensures that $ \tilde{O} $ avoids zero eigenvalues, preserving invertibility while maintaining the physical properties of the system. The expectation value os then given by
\begin{align}
    \braket{\tilde{O}(T)} = \tr\left(V_{\tilde{O}}(T) U_0(T) \rho U_0^\dagger(T) \tilde{O}\right),
\end{align}
and the original expectation value can be recovered by inverting the linear transformation:
\begin{align}
   \braket{O(T)} = \frac{\braket{\tilde{O}(T)} - b}{a}. 
   \label{equ:trnsfm}
\end{align}
This approach ensures that the graybox framework can handle non-invertible observables robustly.

\subsubsection{Choice of basis}

The graybox framework requires an informationally-complete basis for accurately estimating then noise operators. Two commonly used bases for this purpose are the clock-shift basis and the Gell-Mann basis, both of which generalize the Pauli matrices to higher-dimensional systems. The Gell-Mann basis provides a traceless, Hermitian representation. The $ d^2 - 1 $ Gell-Mann matrices are divided into three categories:

1. Symmetric Gell-Mann Matrices:
   $$
   \lambda_{jk}^{\text{(S)}} = \left( \ket{j}\bra{k} + \ket{k}\bra{j} \right), \quad 1\le j \le k \le d.
   $$
2. Anti-symmetric Gell-Mann Matrices:
   $$
   \lambda_{jk}^{\text{(A)}} = -i\left( \ket{j}\bra{k} - \ket{k}\bra{j} \right), \quad 1 \le j \le k \le d.
   $$
3. Diagonal Gell-Mann Matrices:

\begin{multline*}
   \lambda_l = \sqrt{\frac{2}{l(l+1)}} \bigg( \sum_{m=1}^l \ket{m}\bra{m} \\
   - l \ket{l+1}\bra{l+1} \bigg), \quad l = 1, \dots, d-1.
\end{multline*}
Together with the identity operator $ I_N $, these matrices form a complete Hermitian basis for all $ d \times d $ operators. The scaling of all observables in this case grows as $ O(d^5) $, (there are $(d^2-1)$ basis, each has $d$ eigenstates, and we have $d^2$ measurement operators). This imposes a challenge for experimental feasibility as $d$ grows. An alternative basis could be the clock-shift basis presented in  Supplementary Note 1.

\subsubsection{Generalization power of the graybox}
The graybox framework provides a powerful tool for predicting the dynamics of arbitrary observables and initial states in quantum systems. Let the measurement operator $ O $ be expressed as a linear combination of an orthonormal basis (ONB) $ \{A_i\} $, such that $ O = \sum_i a_i A_i $, and let the initial state $ \rho = \sum_j b_j A_j $. The expectation value of the observable $ O$ at time $ T $ is then given by:
\begin{align}
\braket{O(T)}= \sum_{i,j,k} a_i b_j \lambda_j^{(k)} \mathcal{G}(i,j,k),
\end{align}
where $ \mathcal{G}(i,j,k) $ encodes the output of the graybox for the observable $ A_i $, given an initial state corresponding to the $ k^{\text{th}} $ eigenvector of $ A_j $. Therefore, if the graybox framework accurately predicts $ \mathcal{G}(i,j,k) $, it can generalize to compute the expectation values of any observable $ O $ for any initial state $ \rho $. Thus, the graybox framework is not limited to specific observables or states but can fully characterize the dynamics of the system under study. Detailed derivations are provided in Supplementary Note 2.

\subsection{Machine learning architecture}
\begin{figure*}[htbp!]
    \centering
    \includegraphics[width=1.0\textwidth]{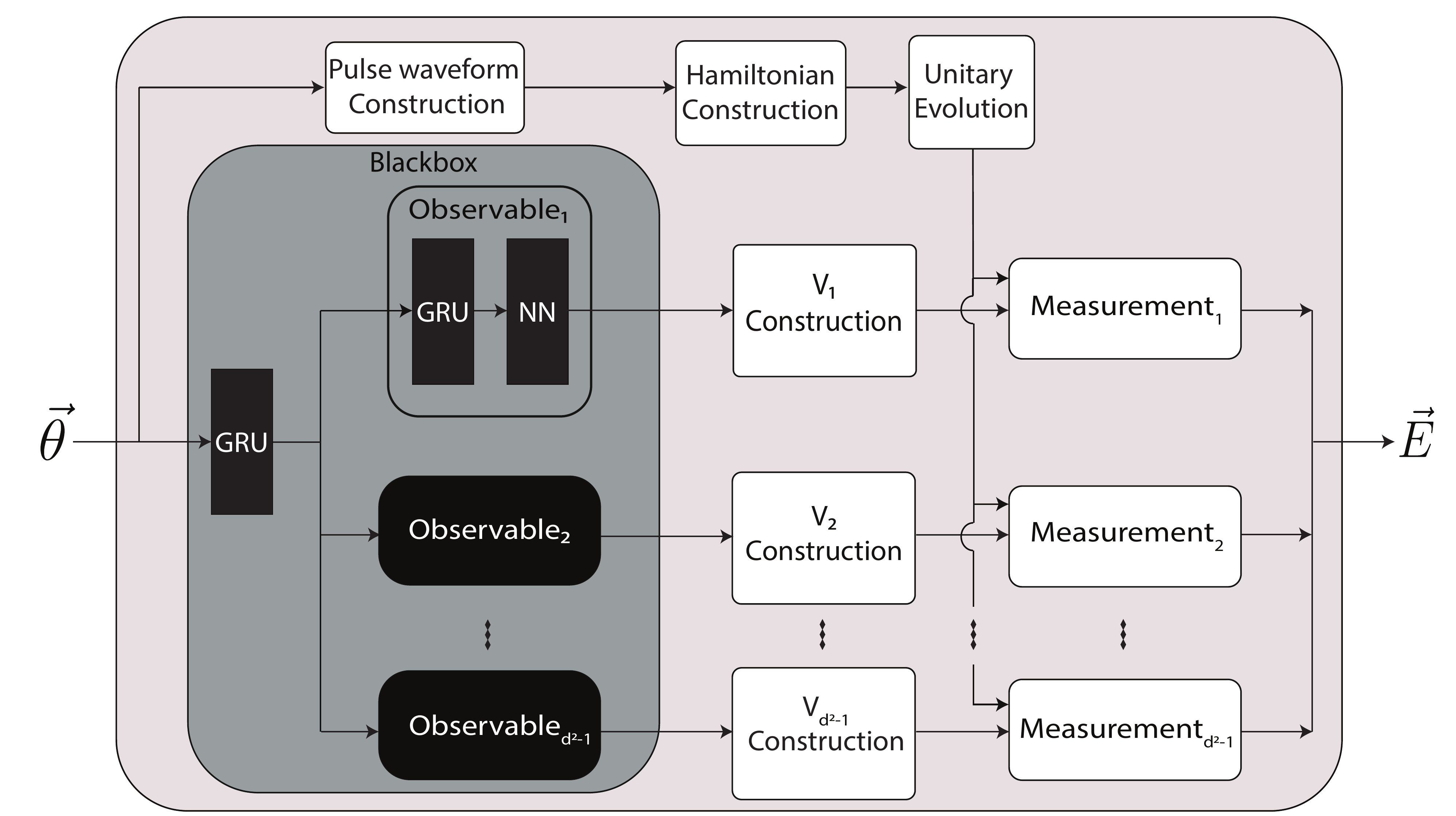}
    \caption{\textbf{Graybox machine learning architecture for qudit identification and control.} The model accepts pulse parameters $\vec{\theta}$ as input and processes them through two parallel components: a whitebox stream that constructs the time-domain control waveform and models Hamiltonian evolution, and a blackbox stream that captures the effective action of the noise. The whitebox component construct the control waveform given the input parameters, then computes the corresponding Hamiltonian, and finally computes the time-ordered unitary evolution under the specified system parameters. In parallel, the blackbox component receives the same pulse parameters and encodes them through an initial GRU layer with 60 hidden units, which projects the input into $d^2 - 1$ streams corresponding to the generalized Gell-Mann measurement basis. Each channel is then processed by an independent GRU  layer with 60 hidden units followed by a neural network layer consisting of $3d(d-1)/2$ nodes with sigmoid activation, and $d$ nodes with hyperbolic tangent, and $d$ with softmax activation. The output of this neural network layers are the parameters (eigenvlaues and eigenvectors) needed to construct the noise operator $V_O$ associated with that observable. The outputs from both the blackbox and whitebox components are merged in a measurement unit that computes expectation values of each operator with respect to a complete measurement basis set of input states. For a qudit of dimension $d$, each of the $d^2 - 1$ observables contributes $d$ eigenstates used as initial states, resulting in a total of $d(d^2 - 1)$ expectation values, $\vec{E}$. }
    \label{fig: Method}
\end{figure*}

\paragraph*{Model:} In the graybox framework, the model predicts the expectation values of observables for a given the control parameters $ \vec{\theta} $, combining a physics-driven formalism with a neural network approximation. Specifically, the graybox model uses a blackbox to approximate the operator $ V_O(T; \vec{\theta}) $. In the following, we describe the model architecture, describing the the physics models (whitebox) and machine learning (blackbox) components in this graybox framework. Figure \ref{fig: Method}) shows the structure of the proposed graybox.

\paragraph*{Blackbox layers:} The exact calculation of the measurement outcomes depends on both the noise and the control pulse sequence. To avoid making specific assumptions about these factors, we employ standard machine learning Blackbox layers, such as neural networks (NN) and gated recurrent units (GRU). These Blackbox layers offer an abstract, assumption-free representation of the noise and its interaction with the control pulses. By using these layers, we can estimate the parameters required for reconstructing the $V_O$ operators within a whitebox framework. The strength of such layers lies in their ability to model complex, non-linear relationships, enabling them to represent unknown maps effectively. Our model incorporates three such layers, which are explained as follows.

\begin{itemize}[leftmargin=1.2em]
  \item \textbf{Initial GRU:} This layer processes the control pulse parameters, performing an initial feature transformation to enhance the learning process. Instead of separate preprocessing, the model learns the optimal transformation within the architecture. The inspiration behind this layer is to provides an abstract representation that corresponds to the interaction unitary, which depends on the noise and control pulses. The input represents the control pulses, the output represents the interaction evolution operator, and the weights represent the noise. The GRU is used to capture temporal dependencies in the sequence, which is essential for modeling relations similar to those in open quantum system dynamics. 

  \item \textbf{Observable GRU:} This GRU layer increases the complexity of the model to capture more intricate relationships. The motivation behind this layer is to have an equivalent representation of the operator $ W_O$. For each measurement operator used, we have an instance of this layer, making a total of $d^2-1$, with 60 hidden nodes per layer. Each of those layers has the input connected to the output of the initial GRU. 

  \item \textbf{Parameters neural network:} A fully connected neural layer with $(3d(d-1)/2) + 2d $ nodes follows the each of the observable GRU layers. The first $3d(d-1)/2$ nodes use a sigmoid activation with range $[0,1]$, to represent the parameters $r$, $\Theta$, and $\Psi$ for each $U_{pq}$ needed to construct the matrix $Q$ in Equation \ref{equ:Q}. The next $d$ nodes are of hyberbolic tanget activation with range $[-1,1]$, to represent the variables $x_i$ in Equation \ref{equ: D}. Finally, the last $d$ nodes have softmax activation to generate the parameters $p_i$ in Equation \ref{equ: D} with the desired properties of a discrete probability distribution. 
\end{itemize}

\paragraph*{Whitebox layers:}
The whitebox components implement the known quantum mechanical relations, ensuring that the model adheres to physical principles. These layers simulate the quantum system's evolution and measurement processes based on the learned parameters. The following whitebox components are used in the model:
\begin{itemize}[leftmargin=1.2em]
    \item \textbf{Pulse waveform generation}: This layer takes the pulse parameters and converts it to the time-domain waveform of the control pulses, discretized into $M$ time steps. Each time step corresponds to the pulse amplitude at a specific moment in time. 
 The time-domain waveform is processed by quantum-specific (whitebox) layers that simulate the quantum evolution.

    \item \textbf{Hamiltonian construction}: This layer takes as input the time-domain representation of the control pulse sequence and constructs the closed-system Hamiltonian $ H_0(t) $, at each time step. 
    \item \textbf{Quantum Evolution}: The layer takes the constructed Hamiltonian and computes the time-ordered evolution to compute the unitary evolution operator, $ U_0(T)$. The evolution is approximated using a product of exponentials over discrete time steps,
\begin{align}        
        U_0(T) \approx \prod_{k=0}^{M-1} \exp\left(-i H_0(t_k) \Delta t \right),  
\end{align}
  where  $\Delta t = \frac{T}{M}$  and $t_k = k \Delta t $ for each time step.
  
    \item \textbf{Noise operator construction}: This layer constructs the noise operator $ V_O(T) $ from the parameters $ r $, $ \Theta $, $ \Psi $, $x_i$, and $ p_i $ generated by the Blackbox layers. In other words, it implements directly Equations \ref{equ:Q}, \ref{equ: D}, followed by Equation \ref{equ: Voparam2} and finally Equation \ref{equ: Voparam1}. For each observable $O$ we construct one such layer.

    \item \textbf{Quantum measurement}: This layer calculates the expectation values of the quantum observables for given an initial state. It takes the noise operator $ V_O(T) $ and the unitary $ U_0(T) $ as input from the other whitebox layers, and generates the expectation of the observable using equation \ref{equ:Vo}. The layer also implements the transformation in Equation \ref{equ:trnsfm} if the observable is non-invertible. The calculations are then repeated for each pair of observable and initial state to generate the output of the model $\vec{E}$, of $d(d^2-1)^2$  elements.

\end{itemize}
\subsection{Interpretable and explainable framework}

The exact calculation of the functional dependence of the noise operator $V_O(T)$ on the control pulses is very hard to capture, especially for continuous waveforms (as opposed to unrealistic instantaneous pulses). Approximation methods such as Dyson expansion \cite{auza2024quantum} could give only reveal limited information about this dependence. Moreover, even this can become challenging in more general cases (such as strong noise), where the expressions become 
analytically intractable. The problem becomes even more sever for high-dimensional quantum systems, where the exponential growth of the Hilbert space makes direct calculations more challenging. This is the main inspiration behind the structure of our graybox, where the evaluation of the noise operator is possible and efficient, while the functional dependence on control remains latent. This naturally injects interpretability to our model, as compared to a fully blackbox (i.e. the absence of all whitebox), in which case only the model outputs representing the expectation values of observables can be evaluated. 

To take this a step further, we introduce a local expansion approach that leverages machine learning and Taylor series approximations to model the effect of control on the system dynamics. Moreover, we pair local expansion (interpretability) with a global cost-landscape (explainable) that visualizes how each control reacts to perturbations, making the system’s sensitivities immediately visible. It also allows us to understand why certain pulses may succeed or fail for different target gates.

\begin{figure*}[htbp!]
    \centering
    \includegraphics[width=0.66\textwidth]{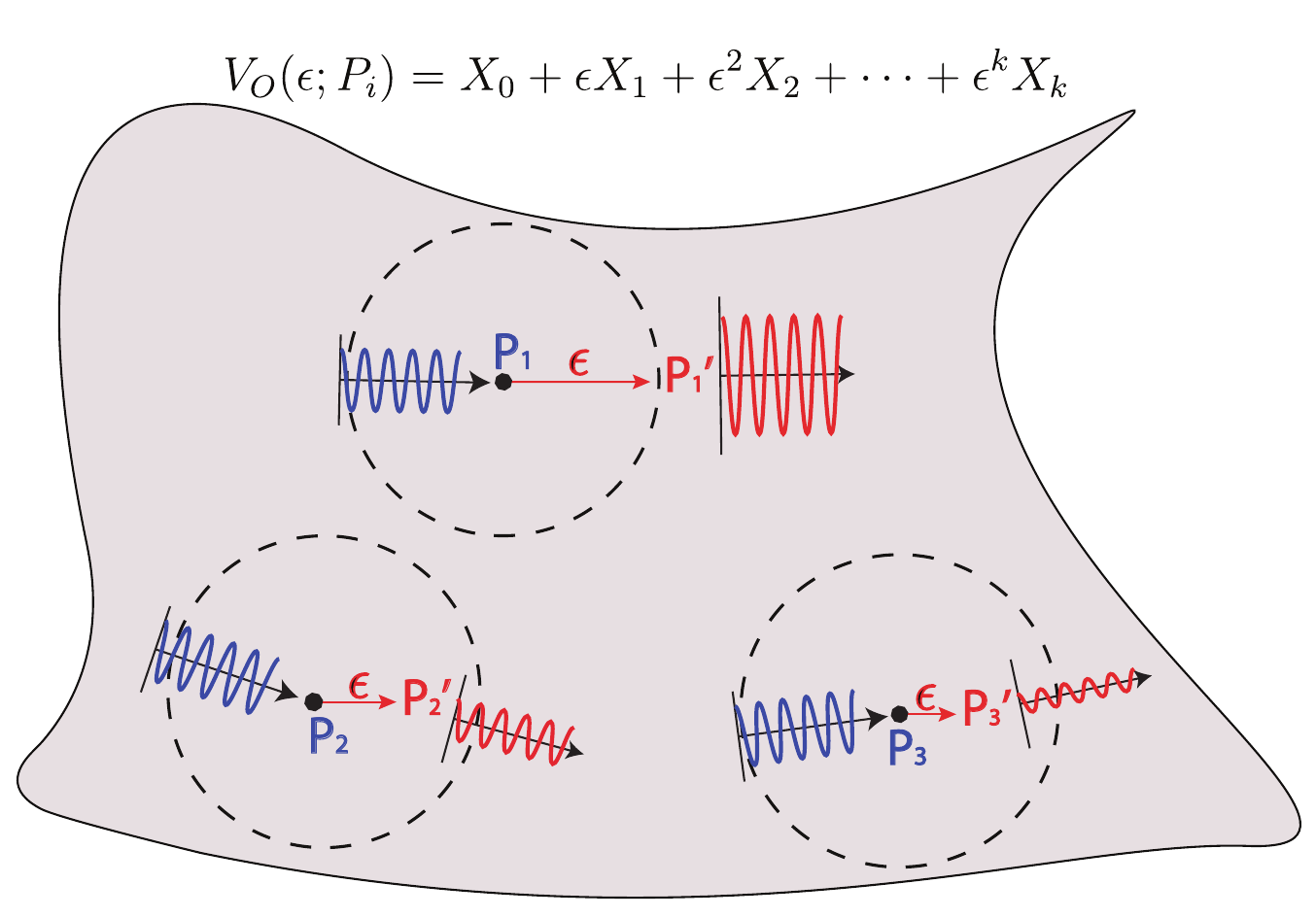}
    \caption{\textbf{Interpretable Learning Framework for Quantum System Identification and Control.} A schematic depicting the interpretability component of the proposed method. In the space of control pulses, from which the training dataset was sampled, the noise operator $V_O(T)$ is expanded locally via a truncated Taylor-like expansion in the neighbourhood of a given control pulse $P_i$, in terms of a small variation $\epsilon$ of the parameters that shifts $P_i$ to $P_i'$. This gives an analytic expression of $V_O(T)$ which can then be used to study how control variations influence the effect of noise on the dynamics of the system, enabling applications in system identification and quantum control design. The Taylor matrix coefficients $X_k$ are estimated by fitting the trained graybox predictions of the matrix elements of $V_O$ to a polynomial expression in $\epsilon$ using points sampled in the neighbourhood under consideration.}
    \label{fig: Method}
\end{figure*}

\subsubsection{Interpretability via local expansions of the noise operator}

In order to construct local expansions of $ V_O(T)$, we first define neighbourhoods around points in the space of control pulses. The choice can be used depending on the application and control pulse shape. Particularly, we look into a small variation $\epsilon$ of the parameters of a pulse $P_i$ that shifts it to $P_i'$ in the space of control pulses. For instance, in the case of Gaussian pulses, the scaling factor $ \epsilon $ could represent variance perturbation, while for square pulses, it could describe changes in pulse width. After defining the neighbourhood in terms of $\epsilon$, we predict $V_O(T)\equiv V_O(\epsilon;P_i)$ from the trained graybox model offline for different values of $ \epsilon $. Next, we fit the predictions to a Taylor-like expansion in the form
\begin{align}
  V_{O}(\epsilon; P_i) = X_0 + \epsilon X_1 + \epsilon^2 X_2 + \cdots+ \epsilon^kX_k,  
  \label{equ: Voexpand}
\end{align}
where the the matrices $ X_k $ describe the $k^{\text{th}}$ Taylor coefficient. These coefficients are obtained by fitting the predictions of the graybox model to a polynomial expression of a given maximum order $k$ given the different values of $ \epsilon $. In particular, the real and imaginary parts of every matrix elements is fitted separately to a polynomial, then the all the $2d^2$ polynomials are then combined again to to find the expression for all $X_k$. The fitted expression of $V_O$ is then checked against the predictions of graybox model to avoid  overfitting or underfitting. This can determine the order to which the Taylor expression is truncated to. Since our dataset spans the full measurement basis, this ensures $V_O(T)$ is interpretable, where for example a noiseless system has $V_O=I$. Figure \ref{fig: Method}) 
depicts the concept of local expansions of the noise operators.

\subsubsection{Explainable (post-hoc interpretable) control cost function}

Using the local Taylor expansion, we define a control cost function $ J(\epsilon; P_i) $ that quantifies the deviation from ideal noiseless evolution:  
\begin{align}
    J(\epsilon; P_i) = \sum_{k=1}^{d^2-1} \| V_{O_k}(\epsilon; P_i) - I \| + \left(1 - \frac{|\mathrm{Tr}(U^\dagger G)|^2}{d^2} \right).
    \label{equ:Intercostfunction}
\end{align}
The first term measures noise strength along a complete basis of operators $O_k$, where values close to zero indicate minimal noise influence and large values indicate strong noise effects. The second term measures infidelity, quantifying how well the applied control pulse implements the target unitary $ G $. The minimum value of this cost function is zero (since both terms are strictly non-negative), which represents the optimal solution where the noise is canceled, and the target unitary is achieved, thus providing a well-motivated metric for quantum control. The closer $ J(\epsilon; P_i) $ is to zero, the better the pulse is at achieving the desired unitary transformation.  

\section{Results}
\subsection{System Model}
In this section, we present the numerical results of our proposed framework for noisy qudit system identification and control. We focus on a truncated anharmonic oscillator system up to dimensions $d$. The total Hamiltonian of the system  $H(t)$, expressed in the computational basis $\{\ket{0}, \ket{1}, \cdots \ket{d-1}\}$, is
\begin{align}
    H(t) = \sum_{j=0}^{d-1} \omega_j |j\rangle \langle j| &+ f(t) (a_d^\dagger + a_d)  \nonumber \\
    &+ \sum_{j=0}^{d-1} g_j \beta_j(t) |j\rangle \langle j|,
    \label{equ: Ham_qudit}
\end{align}
where $\omega_j$ is the natural frequency of the $j^{\text{th}}$ energy level, $f(t)$ is the control pulse sequence, $ a_d^{\dagger}$ and $a_d$  are the creation and annihilation operators truncated to $d$ levels, respectively,  $\beta_j(t)$ is a stochastic process representing the noise along the $j^\text{th}$ energy level, and $g_j$ is the coupling strength. The first term represents the drift of the system, the second term is the control, and the last term is the noise. Notice that this Hamiltonian is written without invoking the rotating-wave approximation. As such, the control field $f(t)$ is expressed as,
\begin{align}
    f(t) =  \sum_{i=0}^{d-2} \Omega_i ( f_{i}^{I}(t) \cos(\omega_{Di} t) + f_{i}^{Q}(t) \sin(\omega_{Di} t))
\end{align}
Where $\Omega_i$ is scaling factor, $f_{i}^{I}(t)$, $f_{i}^{Q}(t)$, are the $i^{\text{th}}$ in-phase and quadrature modulating envelope, respectively and $\omega_{Di}$ is the drive frequency to induce a transition between levels $|i\rangle$ and  $|i+1\rangle$. The individual envelopes are constructed using a Hanning pulse parametrization \cite{theis2016simultaneous, Martinis} of the form,
\begin{align}
    f_i^{(I/Q)}(t) = \sum_{n=1}^{n_{\max}} \frac{1}{2} A_{n,i}^{(I/Q)} \left( 1 - \cos\left( \frac{2\pi n t}{T} \right) \right),
    \label{equ: Hanningenvelope}
\end{align}
where  $A_{n,i}^{(I/Q)}$ is the amplitude of the $n^\text{th}$ pulse for in-phase and quadrature channels along the $i^\text{th}$ direction, and the total number of pulses in the sequence is $n_{\max}$. The Hanning window offers a smooth rise and fall to zero at the endpoints of the interval $[0,T]$. The set of all amplitudes form the control parameter vector $\vec{\theta}$ of $2(d-1)n_{\max} $ elements.

The noise, $\beta_j(t)$ introduced is dephasing non-Markovian noise, with power spectral density $S_\beta(f)$ defined as
\begin{align}
    S_\beta(f)=\frac{\alpha_1}{f}+\alpha_2f,
\end{align}
consisting of both $\frac{1}{f}$ noise and proportional noise components as depicted in Supplementary Figure S1. The $\frac{1}{f}$ noise models low-frequency fluctuations (such as charge noise in superconducting circuits), while proportional noise represents high-frequency fluctuations \cite{superconductingreview}. The different noise processes $\beta_i(t)$ are generated independently. 

\subsection{Dataset generation}
To demonstrate the developed methodology, we selected a qudit of dimension $d=3$ (qutrit), in which case 
\begin{align}
a_3= \sqrt{1}\ket{0} \bra{1} + \sqrt{2}\ket{1} \bra{2}.
\end{align}
In Supplementary Note 3, we also introduce a $d=2$ system, which reduces to a usual qubit. The numerical values for the model parameters are shown in Supplementary Tables S1 and S2 for qubit and qutrit respectively. 

We study three cases: closed system, weak and strong noise coupling regimes to explore the performance of our approach under different conditions. For each setting, we generated a dataset for training the graybox via MATLAB by simulating the time evolution of a qutrit system under the aforementioned system model. A Monte Carlo simulation was performed to approximate the non-Markovian dynamics. The system is initialized in the eigenvectors of the generalized Gell-Mann matrices, and expectation values of the Gell-Mann operators are computed. This process is repeated across varying initial control sequences randomly (by varying the amplitudes of the pulses), to generate the required dataset. In particular, the control fields $ f_1(t) $ and $ f_2(t) $, are generated according to Equation \ref{equ: Hanningenvelope} where $ A_n^{(i)} $ are randomized coefficients sampled uniformly in the range $[-A_{\text{max}}, A_{\text{max}}]$, with $ A_{\text{max}} = \frac{2\pi}{T} $ ensuring appropriate amplitude scaling (for derivation of maximum amplitude, see Supplementary Note 4). The pulse number, $ n_{\max} $, determines the number of Hanning components included, and we used $ n_{\max} = 10 $. The final datasets consisted of $10,032$ examples.

\subsection{Machine learning model training }
The Graybox model is trained via Python machine learning library TensorFlow Keras \cite{tensorflow} on the generated dataset for $10^3$ training iterations. After training, the model is evaluated on a testing dataset, which is separate from the training data. The dataset split was 8192 examples for training, and 1840 for testing. The model's accuracy was assessed using the mean squared error (MSE) between predicted and actual observables. The training results for open (strong noise) and closed system are shown in Supplementary Figure S2. As shown, for the closed system, the training and testing MSE fall to orders of $10^{-5}$ and for the noisy cases, they fall to $10^{-3}$.

\subsection{Qudit system identification}
To demonstrate the interpretability of our framework, we perform a second-order Taylor expansion (Eq. \ref{equ: Voexpand}) of the predicted operator $V_O(\epsilon;P_i)$ from the graybox model. For each control input, the model predicts eight matrix-valued operators $\{ V_{O_1}, \cdots, V_{O_8} \}$, corresponding to the Gell-Mann operators. We construct the expansion for a set of input pulses by scaling the pulse amplitude as $f_\epsilon(t) = \epsilon f(t)$. To illustrate this process, we present the fitted expansion for the operator $V_{O_1}(\epsilon,P_i)$ associated with the pulse indexed $1595$ in the testing dataset (All coefficients are listed in Supplementary Note 5). 

As such, the expansion is given by
\begin{equation}
    V_{O_1}(\epsilon; P_{1595}) = X_0 + \epsilon X_1 + \epsilon^2 X_2
\end{equation}
where for the \emph{strong noise},
\begin{align*}
X_0 &=
\footnotesize
\begin{bmatrix}
0.876-0.006i & 0.060-0.003i & 0.001+0.007i\\
0.063+0.003i & 0.878+0.006i & -0.001-0.008i\\
0.000+0.009i & 0.002-0.006i & 0.999+0.000i
\end{bmatrix},
\\
X_1 &=
\footnotesize
\begin{bmatrix}
-0.012+0.002i & -0.005+0.001i & -0.000-0.001i\\
0.022-0.001i & 0.002-0.002i & 0.000+0.001i\\
0.000-0.001i & -0.000+0.001i & -0.027+0.000i
\end{bmatrix},
\\
X_2 &=
\footnotesize
\begin{bmatrix}
-0.002-0.002i & 0.003-0.001i & 0.000+0.000i\\
-0.002+0.001i & -0.005+0.002i & -0.000-0.000i\\
-0.000+0.000i & 0.000-0.000i & -0.003-0.000i
\end{bmatrix};
\end{align*}
while for the \emph{closed system},
\begin{align*}
X_0 &=
\footnotesize
\begin{bmatrix}
1.000+0.006i & 0.001+0.003i & 0.004-0.010i\\
-0.004-0.003i& 0.998-0.006i & -0.004+0.011i\\
-0.005-0.013i& 0.004+0.008i & 0.997+0.000i
\end{bmatrix},
\\
X_1 &=
\footnotesize
\begin{bmatrix}
0.005-0.001i & -0.003-0.001i & -0.001+0.002i\\
-0.000+0.001i & 0.006+0.001i & 0.001-0.002i\\
0.001+0.002i & -0.001-0.001i & 0.002-0.000i
\end{bmatrix},
\\
X_2 &= 
\footnotesize
\begin{bmatrix}
-0.001-0.003i & -0.003-0.002i & 0.000-0.000i\\
0.001+0.002i & 0.000+0.003i & -0.000+0.000i\\
-0.000-0.001i & 0.000+0.000i & 0.000-0.000i
\end{bmatrix}.
\end{align*}

\subsection{Quantum control}
\subsubsection{Gate optimization}
To asses the performance of our proposed method, we find the optimal pulses to implement a universal set of qutrit gates utilizing the cost function shown as equation \ref{equ:msecostfunction}). We demonstrate the results for two families of gates: Global qutrit gates represented in the clock–shift basis Table \ref{tab:globalresults} defined as:
\begin{align}
 \sigma_{j,k} = \Sigma_1^j \Sigma_3^k,   
\end{align}
where $\Sigma_1^j \Sigma_3^k$, are shift and clock matrices defined in Supplementary Note 1. The second family consists of universal sets defined over the three possible 2-level subspace of the qutrit as shwon in Table \ref{tab:subspaceresults}. Particularly we look at Pauli, Hadamard, and $\pi/4$ $X$-rotation gate.
As shown tables above, the global gates produced infidelities under $0.08$ in strong noise settings for all gates, demonstrating reliable generalization across the ${SU}(3)$ Hilbert space. Furthermore, the sublevel gates produced infidelities of $0.085$ under strong noise conditions for all gates. This confirms that the method exhibits high performance across multiple gate types and noise levels (See Supplementary Table S3 for qubit results).
\FloatBarrier
\begin{table}[ht]
\centering
\caption{Results of graybox optimized qutrit global logic gates in the clock‐shift bases, displaying closed‐system (ideal), weak noise and strong noise infidelities.}
\label{tab:globalresults}
\begin{tabular}{|c|c|c|c|}
\hline
\multicolumn{4}{|c|}{\bfseries Clock-shift (global) gates} \\
\hline

\hline
Operator & Closed system & Weak noise & Strong noise \\
\hline
$\sigma_{0,0}$ & $1.14\times10^{-5}$ & $1.49\times10^{-2}$ & $7.98\times10^{-2}$ \\
$\sigma_{1,0}$ & $5.00\times10^{-6}$ & $1.46\times10^{-2}$ & $6.38\times10^{-2}$ \\
$\sigma_{2,0}$ & $5.35\times10^{-6}$ & $1.52\times10^{-2}$ & $6.44\times10^{-2}$ \\
$\sigma_{0,1}$ & $5.33\times10^{-6}$ & $1.12\times10^{-2}$ & $5.19\times10^{-2}$ \\
$\sigma_{0,2}$ & $3.49\times10^{-5}$ & $1.33\times10^{-2}$ & $5.46\times10^{-2}$ \\
$\sigma_{1,1}$ & $4.35\times10^{-5}$ & $1.31\times10^{-2}$ & $5.15\times10^{-2}$ \\
$\sigma_{1,2}$ & $5.35\times10^{-6}$ & $1.27\times10^{-2}$ & $5.09\times10^{-2}$ \\
$\sigma_{2,1}$ & $3.74\times10^{-6}$ & $1.13\times10^{-2}$ & $4.85\times10^{-2}$ \\
$\sigma_{2,2}$ & $4.21\times10^{-6}$ & $1.12\times10^{-2}$ & $4.97\times10^{-2}$ \\
\hline
\end{tabular}
\end{table}

\begin{table}[ht]
\centering
\caption{Results of graybox optimized qutrit subspace logic gates in the Gell-Mann bases, displaying closed‐system (ideal), weak noise and strong noise infidelities.}
\label{tab:subspaceresults}
\begin{tabular}{|c|c|c|c|c|}
\hline
\multicolumn{5}{|c|}{\bfseries Gell-Mann (subspace) gates} \\
\hline

\hline
\multicolumn{2}{|c|}{Operator} & Closed system & Weak noise & Strong noise \\
\hline
\multirow{5}{0.3cm}{\rotatebox{90}{$\ket{0},\ket{1}$}}
&$X_{01}$ & $2.24\times10^{-5}$ & $1.53\times10^{-2}$ & $7.44\times10^{-2}$ \\
&$Y_{01}$ & $4.35\times10^{-6}$ & $7.72\times10^{-3}$ & $2.95\times10^{-2}$ \\
&$Z_{01}$ & $1.48\times10^{-5}$ & $1.36\times10^{-2}$ & $8.42\times10^{-2}$ \\
&$H_{01}$ & $4.44\times10^{-6}$ & $1.52\times10^{-2}$ & $8.35\times10^{-2}$ \\
&$R_{01}$ & $2.44\times10^{-6}$ & $1.66\times10^{-2}$ & $5.91\times10^{-2}$ \\
\hline
\multirow{5}{0.3cm}{\rotatebox{90}{$\ket{1},\ket{2}$}}
&$X_{12}$ & $6.06\times10^{-6}$ & $1.31\times10^{-2}$ & $6.95\times10^{-2}$ \\
&$Y_{12}$ & $7.40\times10^{-6}$ & $2.00\times10^{-2}$ & $4.75\times10^{-2}$ \\
&$Z_{12}$ & $4.60\times10^{-6}$ & $1.70\times10^{-2}$ & $8.25\times10^{-2}$ \\
&$H_{12}$ & $1.48\times10^{-5}$ & $1.73\times10^{-2}$ & $7.68\times10^{-2}$ \\
&$R_{12}$ & $4.16\times10^{-6}$ & $1.73\times10^{-2}$ & $7.28\times10^{-2}$ \\
\hline
\multirow{5}{0.3cm}{\rotatebox{90}{$\ket{0},\ket{2}$}}
&$X_{02}$ & $6.13\times10^{-6}$ & $1.66\times10^{-2}$ & $7.47\times10^{-2}$ \\
&$Y_{02}$ & $9.11\times10^{-6}$ & $1.22\times10^{-2}$ & $5.35\times10^{-2}$ \\
&$Z_{02}$ & $5.28\times10^{-6}$ & $2.00\times10^{-2}$ & $8.30\times10^{-2}$ \\
&$H_{02}$ & $5.51\times10^{-5}$ & $1.92\times10^{-2}$ & $7.91\times10^{-2}$ \\
&$R_{02}$ & $5.97\times10^{-6}$ & $1.67\times10^{-2}$ & $7.73\times10^{-2}$ \\
\hline
\end{tabular}
\end{table}
\FloatBarrier

\begin{figure*}[htbp!]
    \centering
    \includegraphics[width=1.0\textwidth]{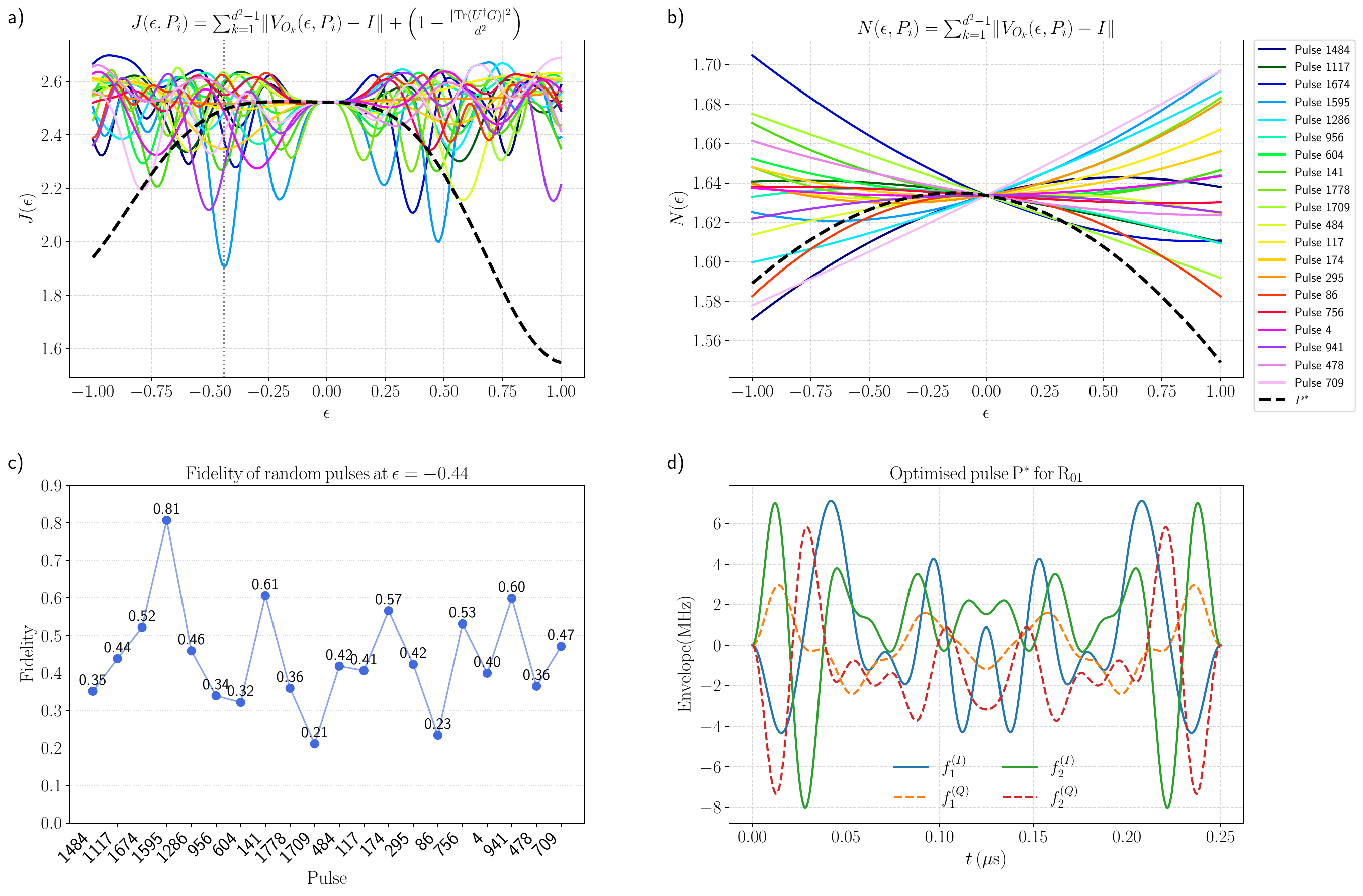}
    \caption{\textbf{Interpretable pulse parameter robustness analysis of qutrit control pulses for implementing target gate, $G=R_{01}$.} (a) Evaluation of the full interpretable control cost function $J(\epsilon, P_i)$, which combines fidelity loss and deviation from the identity in observable space, across 20 randomly sampled control pulses (coloured curves). The graybox optimized pulse $P^*$ (dashed black) achieves consistently lower cost across the perturbation range. The vertical dashed line at $\epsilon = -0.44$ indicates the perturbation point, $\epsilon$ that minimizes the control cost metric, used for fidelity comparison in panel (c). (b) Noise sensitivity analysis using cost function $N(\epsilon, P_i)$, which isolates observable mismatch from gate infidelity. The optimized pulse again shows markedly reduced sensitivity relative to the random set, underscoring its resilience under noise. (c) Fidelity values of each random pulse evaluated at $\epsilon = -0.44$. While a few pulses achieve moderate fidelity (e.g., pulse 1595), the majority exhibit suboptimal performance, demonstrating the difficulty of achieving robust control without guided optimization. (d) Time-domain envelope of $P^*$ —the graybox optimized pulse across the I/Q quadratures for both qutrit transitions.}
    \label{fig: controlsample}
\end{figure*}

\subsubsection{Explainable control cost function}
To understand the performance of pulse the optimization, we evaluate their robustness (robust against  variations in amplitudes ) using Equation \ref{equ:Intercostfunction}, as an alternative physics-based cost function. Figure \ref{fig: controlsample}(a–c) compare random pulses against the graybox-optimized pulse $P^*$ for target gate $R_{01}$. Specifically, Figure \ref{fig: controlsample}(a) depicts the cost function $J(\epsilon)$ for $20$ random pulses (obtained from dataset) revealing variability the cost function, with clear minima at $\epsilon = -0.44$. The graybox optimized pulse (dashed black) outperforms all random pulses. In addition, figure \ref{fig: controlsample}b shows The noise sensitivity metric $N(\epsilon) = \sum_i \|V_{O_k} - I\|$ focusing on the noise cancellation aspect of the optimization problem. Figure \ref{fig: controlsample}c shows the fidelity distribution at $\epsilon = -0.44$ showing the range of gate fidelities among random pulses; revealing that pulse indexed $1595$ produces $0.81$ fidelity which exhibits the highest fidelity. Figure \ref{fig: controlsample}d shows the time-domain waveform of optimal pulse $P^*$, plotted across the four I/Q control channels.

\begin{figure*}[htbp!]
    \centering
    \includegraphics[width=1.0\textwidth]{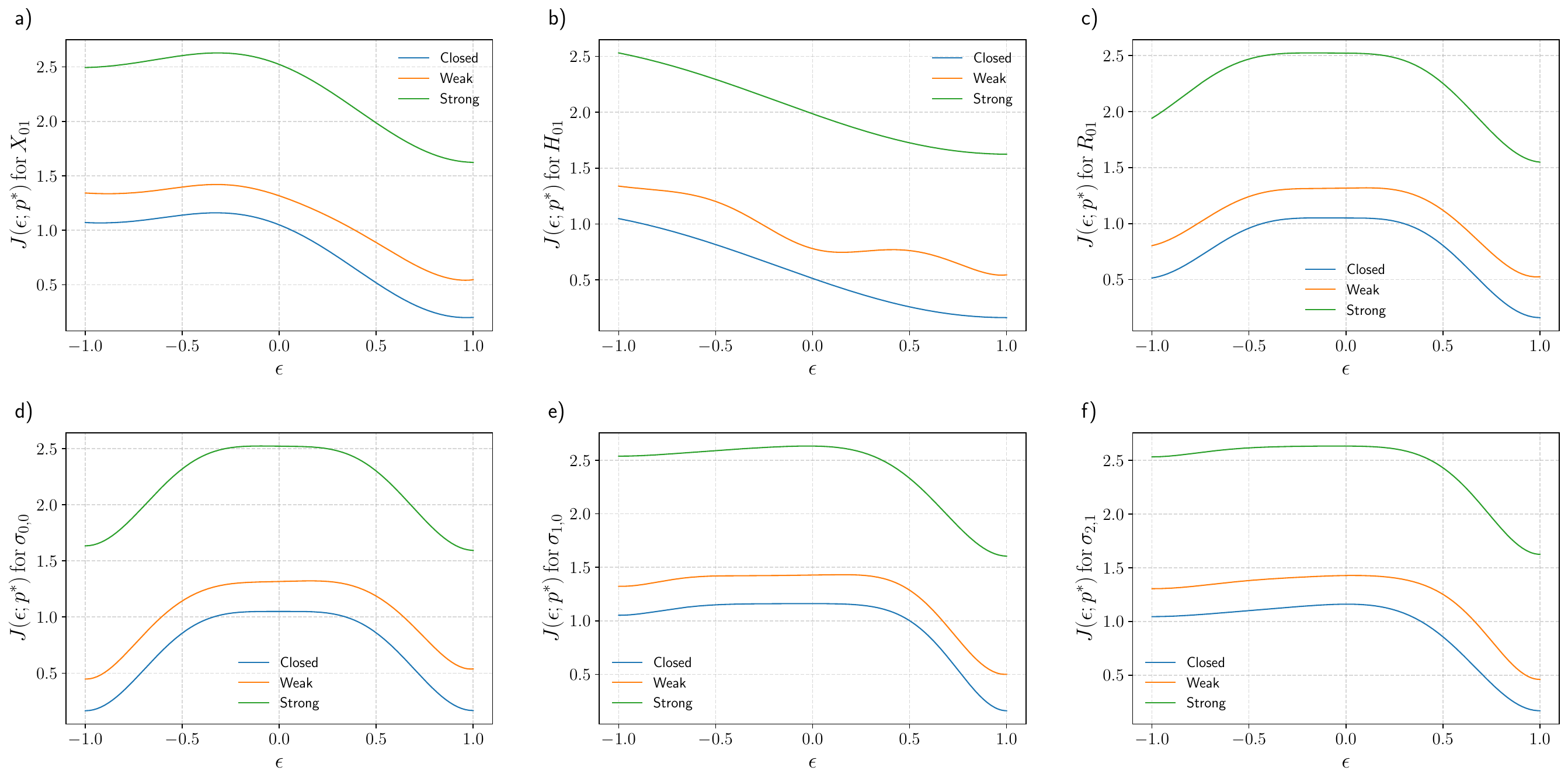}
   \caption{\textbf{Control cost landscapes $J(\epsilon; P^*)$ for six optimized qutrit gates under varying noise conditions.} Each panel depicts the cost function $J(\epsilon)$ as a function of the amplitude scaling parameter $\epsilon \in [-1,1]$, evaluated using a fixed optimized pulse $P^*$ for a specific gate. Curves correspond to three noise regimes: closed system (blue), weak noise (orange), and strong noise (green). \textbf{(a–c)} show results for single-transition Gell-Mann gates $X_{01}$, $H_{01}$, and $R_{01}$; \textbf{(d–f)} display global clock–shift gates $\sigma_{0,0}$, $\sigma_{1,0}$, and $\sigma_{2,1}$, respectively. In all cases, the cost minima occur near $\epsilon = 1$, indicating that the optimizer has matched pulse power to the target gate.}
    \label{fig: controlgates}
\end{figure*}

Figure \ref{fig: controlgates} displays the control cost function $J(\epsilon; P^*)$ evaluated for six optimized qutrit gates across the closed system, weak noise, and strong noise regime. Each subplot corresponds to a distinct target gate $G$, including three subspace gates—$X_{01}$, $H_{01}$, and $R_{01}$—and three global gates—$\sigma_{0,0}$, $\sigma_{1,0}$, and $\sigma_{2,1}$. For all gates, the closed-system curves exhibit consistently low cost near $\epsilon = 1$. Under weak and strong noise, each gate displays a distinct minimum in $J(\epsilon)$, with variation in both the location and depth of the minimum depending on the gate and noise level. The full set of curves is plotted using consistent axis scaling.

\section{Discussion}

In this paper, we proposed a method for ise characterization and control of qudit, bridging the gap between theoretical dynamical models and practical implementation challenges. The proposed approach presents a flexible and scalable alternative to traditional noise spectroscopy methods. As such, this work addresses a critical methodological gap in quantum control, providing a framework that is sufficient to handle the four main challenges outlined in the introduction. First, regarding the challenge pertaining to the breakdown of the RWA, our approach inherently accounts for this providing a general structure independent of the control Hamiltonian, enabling a non-RWA Hamiltonian to be encoded in the white part of the graybox without affecting the results. Second, practical control pulses deviate from idealized impulses. Graybox protocol is independent of the pulse shape as long as it is possible to parametrize efficiently. Finally, for model mismatch and non-Markovian noise challenges, the combination of the data-driven framework with the noise operator formalism makes it adaptable to any setting. This includes model mismatch in a fabricated device, and/or unknown noise environment. Overall, the proposed framework provides a complementary alternative to PSD-based methods including for instance \cite{Alvarez2011}. 

By examining the local expansion of the noise operator for pulse indexed 1595, we see that the closed-system case recovers an operator nearly equal to the identity, with negligible contributions from higher-order terms—serving as expected from the physical meaning of the noise operator. In other words, the control does not affect that $V_O(T)$ as expected. In contrast, the strong-noise case exhibits stronger linear and quadratic contributions, reflecting the system’s sensitivity to noise and control perturbations.

Regarding the performance of the optimized gates, as in Tables \ref{tab:globalresults} and \ref{tab:subspaceresults}, in the closed-system regime, every gate attains near-ideal fidelity (infidelity $< 6\times10^{-5}$). Weak noise raises infidelity in the $10^{-2}$ range, whereas strong noise keeps them below $8.5\times10^{-2}$. Some global clock-shift gates and subspace rotations, particularly $Y_{01}$, $Y_{12}$, and $\sigma_{2,1}$ exhibit the lowest infidelities ($\approx2.95\times10^{-2}$–$4.85\times10^{-2}$), suggesting that some gates might be easier to implement compared to others given a certain noise environment, raising the question of what is an optimal universal gateset in an open quantum system. Notably, $Y_{01}$ posts the best overall figures, with an infidelity of $7.72\times10^{-3}$ under weak noise and $2.95\times10^{-2}$ under strong noise. 

By looking at Figure \ref{fig: controlsample}), we observed that one random pulse achieved a global minimum of the cost function $J(\epsilon)$ at $\epsilon \approx 0.44$. At this same amplitude, it also attained the highest fidelity (0.81), verifying that $J(\epsilon)$ reliably tracks gate performance. Therefore, the cost function landscape accurately singles out the most promising pulses—even within an unoptimized ensemble. The graybox-optimized pulse $P^*$, however, achieves a significantly lower minimum in $J$, further underscoring the effectiveness of the learned control.  The use of the noise sensitivity function in Figure \ref{fig: controlsample}b allows one to assess pulse quality of noise cancelling independent of the specific target gate. The optimized pulse $P^*$ minimizes $N(\epsilon)$ near $\epsilon = 1$ and remains among the least sensitive across the full range of perturbations, indicating that the model has learned to suppress noise pathways whilst compensating for the downstream target. Practically, this allows experimentalists to trade-off fidelity and resilience by tuning the drive amplitude to suit the control context. The framework thus \emph{explains} why the graybox optimizer excels: it locates control pulses that lie near the global noise floor while, also adapting to specific unitary targets. Moreover, the cost function curves of all the pulses intersect at $\epsilon=0$ showing consistency between the different local expansions at this particular point (absence of control). This shows that our approach successfully provides an interpretable framework for quantum system identification and control. 

In Figure \ref{fig: controlgates}), we see that the higher the noise strength is, the higher the cost function is shifted, indicating a harder the system to control as expected. This shows that this cost function provides a more efficient indicator for the control performance compared to calculating process fidelity, while being consistent with it. Finally, while the current analysis fixes $G$ and scans over $\epsilon$, we could also look into how the cost function varies by varying a parameter of the target $G$ (such as the rotation angle). This could help in answering the aforementioned question about the optimal universal gate-set of an open quantum system.

\section{Conclusion and outlook}
In this work, we generalized the graybox machine learning approach to qudits, presenting a framework for qudits capable of learning non-RWA dynamics, non-ideal pulse shapes, and non-Markovian noise. This approach provides an interpretable and explainable data-driven framework for quantum control and noise characterization. Our results highlight the importance of systematic, noise-aware optimization strategies for improving gate fidelity in NISQ devices, where conventional techniques often fail.

Looking forward, this framework could be extended to continuous-variable or infinite-dimensional systems. We could also explore the integration of optimized, interpretable cost functions into reinforcement learning (RL) to enable closed-loop, noise-resilient control. In this setting, the model-based RL component will leverage the generalized noise operator for systematic noise characterization, while the control component will support adaptive pulse design. Another possible extension is to examine ``whitening'' the current graybox model by having a minimal set of assumptions on the noise, where it is known to be satisfied for a given device. Future work will integrate these ideas into enhanced graybox architectures to deliver fully interpretable models for open-qudit dynamics. Together, these contributions advance the development of scalable, high-precision control strategies for next-generation quantum technologies operating in realistic, noisy environments.

\noindent\textbf{Data and code availability} The source code and data generated in this study are available upon  request from the corresponding author.\\

\noindent\textbf{Acknowledgements} AP acknowledges an RMIT University Vice-Chancellor’s Senior Research Fellowship and a Google Faculty Research Award. This work was supported by the Australian Government through the Australian Research Council under the Centre of Excellence scheme (No: CE170100012).  Work at Griffith University was supported (partially) by the Australian Government via AUSMURI Grant (No: AUSMURI000002). This research was also undertaken with the assistance of resources from the National Computational Infrastructure (NCI Australia), an NCRIS enabled capability supported by the Australian Government. \\

\noindent\textbf{Author Contributions}
YM conducted all the qutrit numerical experiments in the paper. RS conducted the single qubit results in the paper. YM and AY designed and implemented the GB architecture. YM, AY, and RS designed and implemented the simulator. GP contributed to the theoretical foundations of the noise operator formalism. AY and AP supervised the project.

\bibliographystyle{apsrev4-2}
%

\end{document}


\title{Supplementary Materials for \\ Quantum Engineering of Qudits with Interpretable Machine Learning}

\author{Yule Mayevsky}
\address{Quantum Photonics Laboratory and Centre for Quantum Computation and Communication Technology, RMIT University, Melbourne, VIC 3000, Australia}

\author{Akram Youssry}
\address{Quantum Photonics Laboratory and Centre for Quantum Computation and Communication Technology, RMIT University, Melbourne, VIC 3000, Australia}
\email{akram.youssry.mohamed@rmit.edu.au}

\author{Ritik Sareen}
\address{Quantum Photonics Laboratory and Centre for Quantum Computation and Communication Technology, RMIT University, Melbourne, VIC 3000, Australia}

\author{Gerardo A. Paz-Silva}
\address{Centre for Quantum Dynamics and  Centre for Quantum Computation and Communication Technology, Griffith University, Brisbane, Queensland 4111, Australia}

\author{Alberto Peruzzo}
\address{Quantum Photonics Laboratory and Centre for Quantum Computation and Communication Technology, RMIT University, Melbourne, VIC 3000, Australia}
\address{Quandela, Massy, France}

\maketitle
\section{Supplementary Note 1: Clock-Shift basis for qudits}

The clock-shift basis 
elements are constructed as:
\begin{align}
 \sigma_{j,k} = \Sigma_1^j \Sigma_3^k,   
\end{align}
where $ \Sigma_1 $ (the ``shift'' matrix) and $ \Sigma_3 $ (the ``clock'' matrix) are defined as:
\begin{align}
\Sigma_1 = \begin{bmatrix}
0 & 0 & \cdots & 1 \\
1 & 0 & \cdots & 0 \\
0 & 1 & \cdots & 0 \\
\vdots & \vdots & \ddots & \vdots
\end{bmatrix}, \quad \Sigma_3 = \begin{bmatrix}
1 & 0 & 0 & \cdots \\
0 & \omega & 0 & \cdots \\
0 & 0 & \omega^2 & \cdots \\
\vdots & \vdots & \vdots & \ddots
\end{bmatrix},
\end{align}
with $ \omega = e^{2\pi i / d} $ being the $ d $-th root of unity. These matrices satisfy:
\begin{align}
\Sigma_1^d = \Sigma_3^d = I, \quad \Sigma_1 \Sigma_3 = \omega \Sigma_3 \Sigma_1.
\end{align}
The set $ \{\sigma_{j,k}\} $ forms a complete basis for all $ d \times d $ operators, with $ j, k \in \{0, 1, \dots, d-1\} $. This basis is invertible however is not Hermitian making them unsuitable for experimental measurements. To address this, a new Hermitian set can be constructed by taking the Hermitian and anti-Hermitian parts of each element. In other words, the basis set becomes

 \begin{align}
     \left\{\frac{1}{2}\left(\sigma_{j,k} + \sigma_{j,k}^{\dagger}\right), \frac{1}{2i}\left(\sigma_{j,k} - \sigma_{j,k}^{\dagger}\right)\right\}, \qquad \forall j,k \in \{0,1,\cdots, d-1\}.
 \end{align}

\section{Supplementary Note 2: Generalisation power of the Graybox (GB)}
\label{subsubsec:GB-generalisation}

 Let the measurement operator $O = \sum_i{a_i A_i}$ for an ONB $\{A_i\}$. Then
 
 \begin{align}
     \braket{O(T)} &=  \sum_i{a_i \braket{A_i(T)}} \\
     &=\sum_i a_i{\tr{(V_{A_i} U_0(T) \rho U_0^{\dagger}(T) A_i)}} \\
 \end{align}
 
 Now, assume we start from the initial state $\rho = \sum_j{b_j A_j}$, and each basis element $A_j$ can be eigendecomposed into $A_j = \sum_k\lambda_k^{(j)} \ket{q_k^{(j)}}\bra{q_k^{(j)}}$. Then
 
 \begin{align}
     \braket{O(T)} &= \sum_i a_i{\tr{(V_{A_i} U_0(T) \rho U_0^{\dagger}(T) A_i)}} \\
     &= \sum_i a_i{\tr{\left(V_{A_i} U_0 (T)\left(\sum_j {b_j A_j} \right) U_0^{\dagger}(T) A_i\right)}}\\
     &= \sum_{i,j} a_i{\tr{ \left(\sum_j {b_j V_{A_i} U_0(T) A_j U_0^{\dagger}(T) A_i}  \right)}} \\
     &= \sum_{i,j}{a_i b_j \tr{ \left(V_{A_i} U_0(T) A_j U_0^{\dagger}(T) A_i \right)}} \\
      &= \sum_{i,j}{a_i b_j \tr{ \left(V_{A_i} U_0(T) \left(\sum_k\lambda_k^{(j)} \ket{q_k^{(j)}}\bra{q_k^{(j)}}\right) U_0^{\dagger}(T) A_i \right)}}\\
      &= \sum_{i,j,k}{a_i b_j \lambda_j^{(k)} \tr{ \left(V_{A_i} U_0(T)\left(\ket{q_k^{(j)}}\bra{q_k^{(j)}}\right) U_0^{\dagger}(T) A_i \right)}}
 \end{align}
 
 Let $\hat{\mathcal{G}}(i,j,k)$ be the output of the GB for observable $A_i$, and initial state being the $k^{\text{th}}$ eigenvector of the $A_j$ operator. Then, we can write the general expectation finally as
 
 \begin{align}
     \braket{O(T)} &= \sum_{i,j,k}{a_i b_j \lambda_j^{(k)} \hat{\mathcal{G}}(i,j,k)}.
 \end{align}
 
 Therefore, if the GB can accurately predict the true expectation $\mathcal{G}(i,j,k)$, then we can predict the observable for any possible initial state/measurement operator.

\section{Supplementary Note 3:Quantum System}
\label{sec:quantumsystem}
\subsection{Qubit System ($  d=2 $)}
\begin{figure*}[htbp!]
    \centering
    \includegraphics[width=0.65\textwidth]{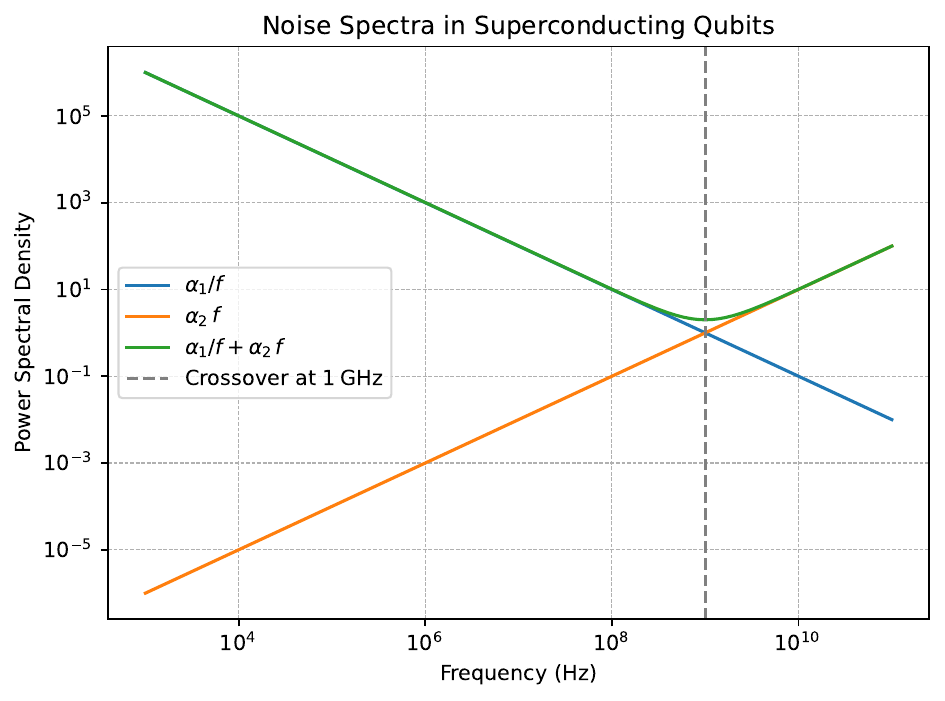}
    \caption{\textbf{PSD Noise Plot} }
    \label{fig: noisepsd}
\end{figure*}

For the qubit system, the Hamiltonian is:

    \begin{align}
            H(t) &= H_{\text{drift}}(t) +H_{\text{control}}(t) + H_{\text{noise}}(t)\\
            &= \frac{1}{2} \omega Z+ \Omega_1  f(t) X +g \beta(t) Z,
    \end{align}

where $ Z $ is the Pauli-$ Z $ operator and $ \omega $ is the natural frequency, $ X $ is the Pauli-$ X $ operator, and $ \Omega_1 $ is the control coupling strength, $ g $ is the noise coupling strength and $\beta(t) $ represents stochastic noise. The noise, $\beta(t)$ is generated from a PSD depicted as show in Figure \ref{fig: noisepsd}
with coefficients chosen so that the two contributions cross over at $ 1 \mathrm{GHz}$. 
\FloatBarrier

\FloatBarrier

  
\FloatBarrier
\begin{table}[h]
\caption{System Parameters for Qubit Dataset Generation}
\label{table:system_parametersqubit}
\centering
\begin{tabular}{|c|c|}
\hline
\textbf{Parameter} & \textbf{Value} \\
\hline
Evolution time $T$ & $0.25\mu s$ \\
Time steps, $M$ & 13250 \\
Realisations $K$ & 3000 \\
Anharmonicity, $\eta$ & $ 200$ MHz \\
Detuning, $\gamma_1$ & $40$ MHz \\
Detuning, $\gamma_2$ & $30$  MHz \\
Carrier amplitude, $\Omega_1$ & 2\\
Oscillation frequency $\omega_1$ & $2\pi \times (5.33$ GHz) \\
Drive frequency $\omega_{D1}$ & $2\pi\times (5.37$ GHz)\\ 
Noise sublevel 1 $g_1$ & 110 (strong)/50 (weak)/ 0 (closed) \\
Number of pulses $n_{\max}$ & 10 \\
Maximum amplitude $A_{\max,H}$ & $25.13$ MHz \\
Low frequency noise amplitude  $\alpha_1$ & $1 \times 10^9$ \\
High frequency noise amplitude $\alpha_2$ & $1 \times 10^{-9}$\\
\hline
\end{tabular}
\end{table}





\begin{table}[h]
\caption{System Parameters for Qutrit Dataset Generation}
\label{table:system_parametersqutrit}
\centering
\begin{tabular}{|c|c|}
\hline
\textbf{Parameter} & \textbf{Value} \\
\hline
Evolution time $T$ & $0.25\mu s$ \\
Time steps, $M$ & 13250 \\
Realisations $K$ & 3000 \\
Anharmonicity, $\eta$ & $ 200$ MHz \\
Detuning, $\gamma_1$ & $40$ MHz \\
Detuning, $\gamma_2$ & $30$  MHz \\
Carrier amplitude, $\Omega_1$ & 2\\
Carrier amplitude, $\Omega_2$ & 2\\
Oscillation frequency $\omega_1$ & $2\pi \times (5.33$ GHz) \\
Oscillation frequency $\omega_2$ & $2\pi \times (10.46 $ GHz) \\
Drive frequency $\omega_{D1}$ & $2\pi\times (5.37$ GHz)\\ Drive frequency $\omega_{D2}$ & $2\pi\times(5.16$ GHz) \\
Noise sublevel 1 $g_1$ & 110 (strong)/50 (weak)/ 0 (closed) \\
Noise sublevel 2 $g_2$ & 120 (strong)/60(weak)/0 (closed)\\
Number of pulses $n_{\max}$ & 10 \\
Maximum amplitude $A_{\max,H}$ & $25.13$ MHz \\
Low frequency noise amplitude  $\alpha_1$ & $1 \times 10^9$ \\
High frequency noise amplitude $\alpha_2$ & $1 \times 10^{-9}$ \\
\hline
\end{tabular}
\end{table}

\FloatBarrier

\section{Supplementary Note 4: Determining the maximum amplitude of the Hanning window control pulse}
\label{sec:maxamp}
To perform a complete $\pi$-rotation the control envelope $f(t)$ must satisfy the pulse‐area condition  
\begin{equation}\label{eq:area}
\int_{0}^{T} f(t)\,\mathrm{d}t \;=\;\pi.
\end{equation}
A convenient choice is to expand $f(t)$ in a finite cosine series of order $d$:  
\begin{equation}
    f(t) =\; \sum_{n=1}^{d} \frac{C_{n}}{2}\Bigl[1 - \cos\frac{2\pi n t}{T} \Bigl].
    \label{eq: hanning}
\end{equation}
Inserting Eq.\ref{eq: hanning} into Eq.\ref{eq:area} gives  
\begin{equation}
    \int_{0}^{T} f(t)\,\mathrm{n_{\text{max}}t
    =\sum_{n=1}^{n_{\text{max}}}\frac{C_n}{2}\int_0^T(1-\cos\frac{2\pi nt}{T})dt=\sum_{n=1}^{n_{{max}}}\frac{C_n}{2}(T-\frac{T\sin(2\pi n)}{2\pi n}) =\sum_{n=1}^{n_{max}}}\frac{C_{n}\,T}{2} =\pi.
    \label{eq: sum_area}
\end{equation}

If we choose all coefficients equal to the same maximum amplitude $A_{\max}$, then Eq \ref{eq: sum_area} implies  
\begin{equation}
    A_{\max} \;=\; \frac{2\pi}{n_{\text{max}}}\,T.
    \label{eq: AmaxH}
\end{equation}
 
The instantaneous envelope satisfies
\begin{equation}
    \bigl\lvert f(t)\bigr\rvert
    \le \sum_{n=1}^{n_{\max}}\frac{\lvert C_n\rvert}{2}\bigl\lvert1-\cos(2\pi n t/T)\bigr\rvert \le \sum_{n=1}^{n_{\max}} \lvert C_n\rvert =n_{\max}\,A_{\max} =\frac{2\pi}{T}.
\end{equation}
By ensuring that both $+\pi$ and $-\pi$ pulses are accessible with the amplitude bound in Eq.\ref{eq: AmaxH}, one can design pulse sequences that compensate residual over‐rotations and recover the desired state.

\begin{figure}[H]
    \centering
    \begin{subfigure}[b]{0.5\textwidth}
        \centering
        \includegraphics[width=\textwidth]{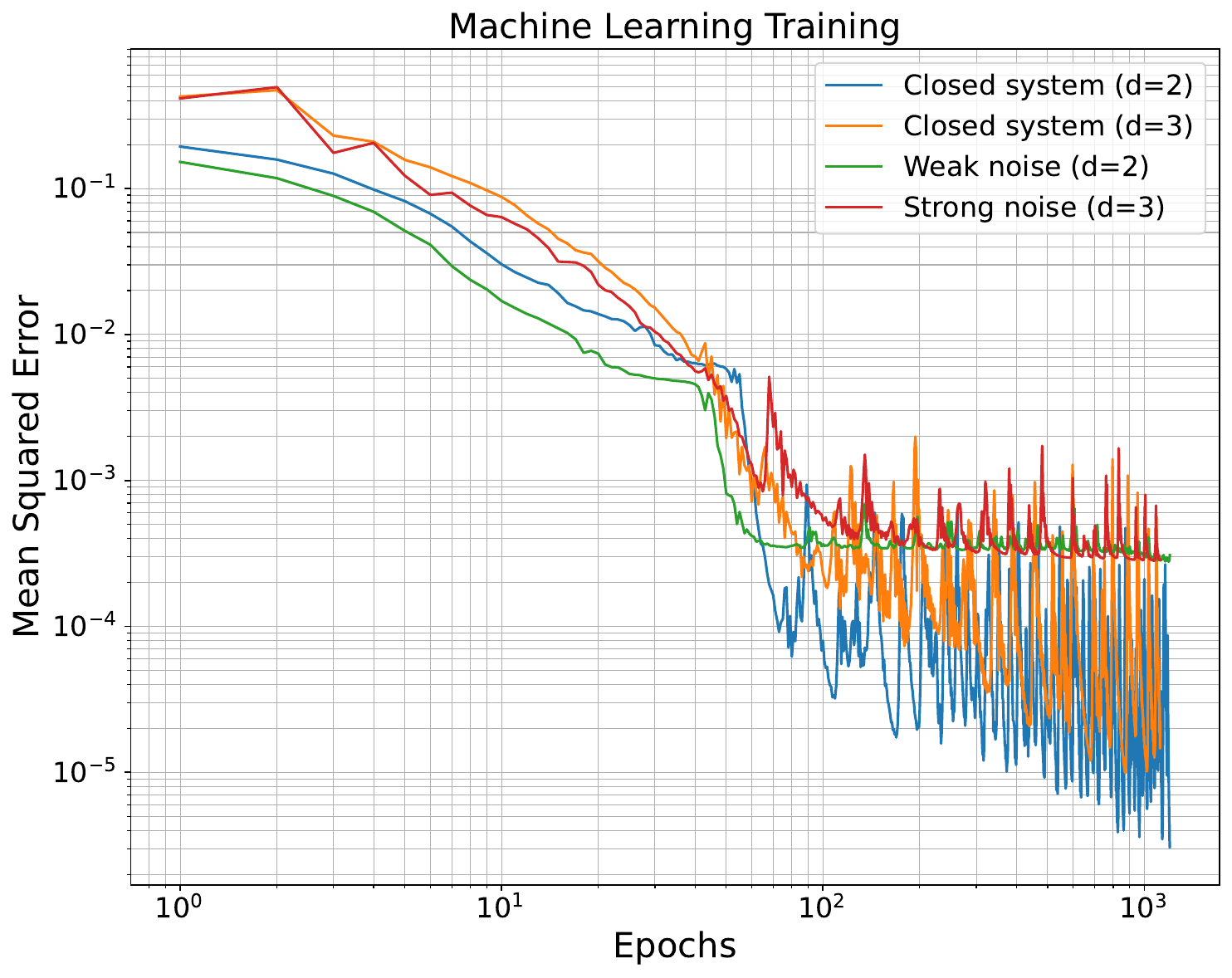}
        \caption{}
        \label{fig:GB tra}
    \end{subfigure}\hfill
    \begin{subfigure}[b]{0.5\textwidth}
        \centering
        \includegraphics[width=\textwidth]{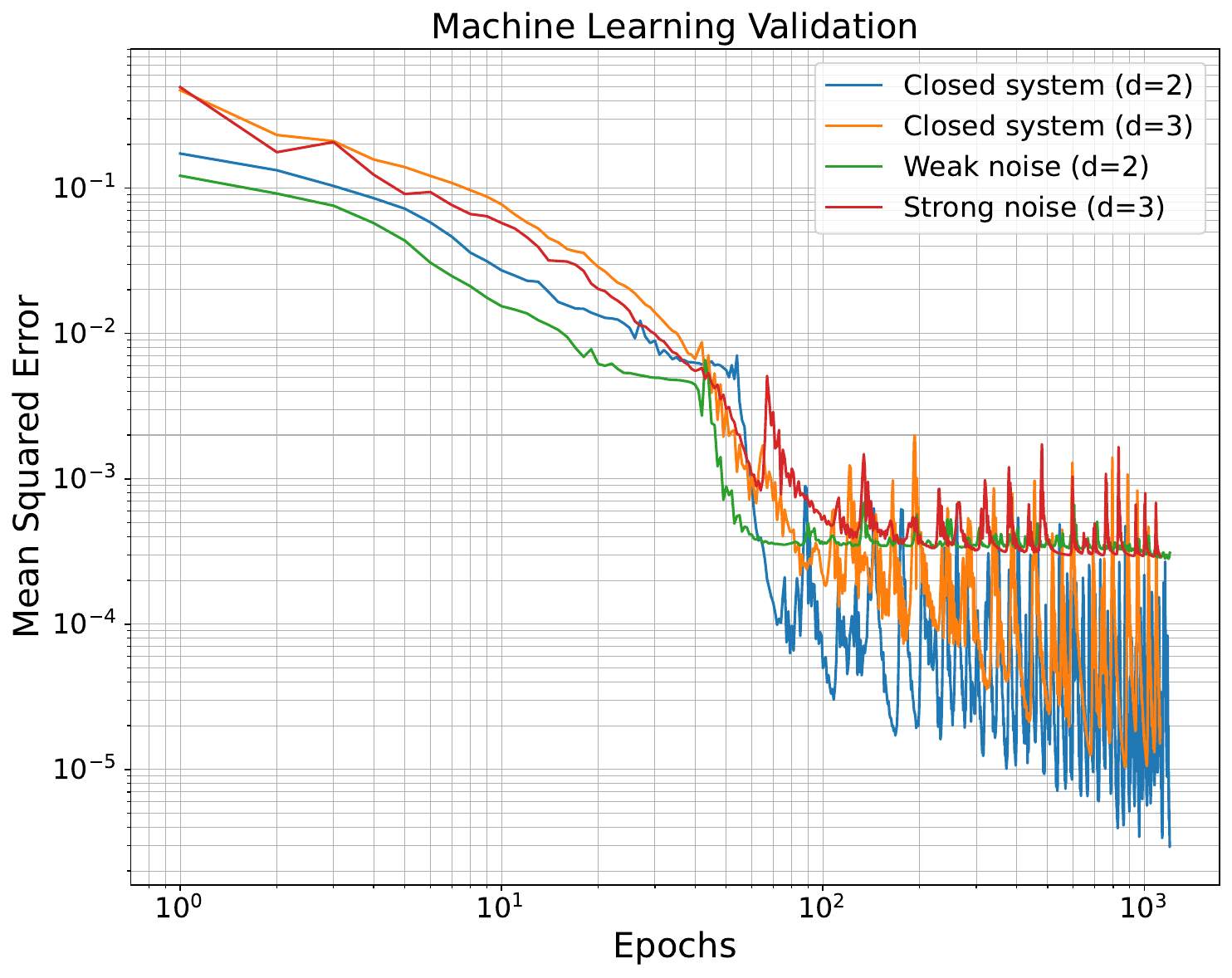}
        \caption{}
        \label{fig:GB val}
    \end{subfigure}\hfill
    \caption{\textbf{The training and testing results of the graybox on qubit and qutrit system.} a) Training Mean Squared Error (MSE) for a closed system setting, weak noise and strong noise. b) Corresponding testing (MSE) curves.}
    \label{fig:evolution_comparison}
\end{figure}

\FloatBarrier  

\begin{table}[ht]
 \centering
 \caption{Infidelities (1 - Fidelity) of qubit gates under different noise conditions.}
 \label{tab:qubit_results}
 \begin{tabular}{|c|c|c|c|}
 \hline
 Gate & Closed system & Weak noise & Strong noise \\
 \hline
 I & $8.0\times10^{-6}$   & $3.38\times10^{-2}$  & $6.87\times10^{-2}$ \\
 X & $5.0\times10^{-6}$   & $2.98\times10^{-2}$  & $5.72\times10^{-2}$ \\
 Y & $1.9\times10^{-5}$   & $2.96\times10^{-2}$  & $5.60\times10^{-2}$ \\
 Z & $4.6\times10^{-5}$   & $3.77\times10^{-2}$  & $6.87\times10^{-2}$ \\
 H & $8.0\times10^{-6}$   & $3.40\times10^{-2}$  & $6.59\times10^{-2}$ \\
 R & $3.9\times10^{-5}$   & $3.30\times10^{-2}$  & $6.64\times10^{-2}$ \\
 \hline
 \end{tabular}
 \end{table}
\section{Supplementary Note 5: Interpretable machine learning results}
\label{sec:intMLsysid}
\begin{equation}
    V_{1}(\epsilon; p_{1595}) = X_0 + \epsilon X_1 + \epsilon^2 X_2
\end{equation}
where for the \emph{strong noise},
\begin{align*}
X_0 &=
\footnotesize
\begin{bmatrix}
0.876-0.006i & 0.060-0.003i & 0.001+0.007i\\
0.063+0.003i & 0.878+0.006i & -0.001-0.008i\\
0.000+0.009i & 0.002-0.006i & 0.999+0.000i\\
\end{bmatrix},\\\\
X_1 &=
\footnotesize
\begin{bmatrix}
-0.012+0.002i & -0.005+0.001i & -0.000-0.001i\\
0.022-0.001i & 0.002-0.002i & 0.000+0.001i\\
0.000-0.001i & -0.000+0.001i & -0.027+0.000i\\
\end{bmatrix},\\\\
X_2 &=
\footnotesize
\begin{bmatrix}
-0.002-0.002i & 0.003-0.001i & 0.000+0.000i\\
-0.002+0.001i & -0.005+0.002i & -0.000-0.000i\\
-0.000+0.000i & 0.000-0.000i & -0.003-0.000i\\
\end{bmatrix};
\end{align*}
while for the \emph{closed system},
\begin{align*}
X_0 &=
\footnotesize
\begin{bmatrix}
1.000+0.006i & 0.001+0.003i & 0.004-0.010i\\
-0.004-0.003i & 0.998-0.006i & -0.004+0.011i\\
-0.005-0.013i & 0.004+0.008i & 0.997+0.000i\\
\end{bmatrix},\\\\
X_1 &=
\footnotesize
\begin{bmatrix}
0.005-0.001i & -0.003-0.001i & -0.001+0.002i\\
-0.000+0.001i & 0.006+0.001i & 0.001-0.002i\\
0.001+0.002i & -0.001-0.001i & 0.002-0.000i\\
\end{bmatrix},\\\\
X_2 &=
\footnotesize
\begin{bmatrix}
-0.001-0.003i & -0.003-0.002i & 0.000-0.000i\\
0.001+0.002i & 0.000+0.003i & -0.000+0.000i\\
-0.000-0.001i & 0.000+0.000i & 0.000-0.000i\\
\end{bmatrix}.
\end{align*}
\begin{equation}
    V_{2}(\epsilon; p_{1595}) = X_0 + \epsilon X_1 + \epsilon^2 X_2
\end{equation}
where for the \emph{strong noise},
\begin{align*}
X_0 &=
\footnotesize
\begin{bmatrix}
0.879-0.005i & -0.002-0.064i & -0.003+0.001i\\
-0.002+0.059i & 0.877+0.005i & 0.002-0.002i\\
-0.007+0.004i & 0.000+0.007i & 0.998+0.000i\\
\end{bmatrix},\\\\
X_1 &=
\footnotesize
\begin{bmatrix}
0.013-0.002i & -0.001-0.024i & 0.000-0.000i\\
-0.001-0.028i & -0.013+0.002i & -0.001+0.000i\\
0.001-0.002i & -0.001-0.001i & 0.012-0.000i\\
\end{bmatrix},\\\\
X_2 &=
\footnotesize
\begin{bmatrix}
-0.005+0.006i & 0.003+0.003i & -0.000+0.000i\\
0.003+0.004i & -0.001-0.006i & 0.000-0.000i\\
-0.000+0.000i & 0.000+0.000i & 0.001-0.000i\\
\end{bmatrix};
\end{align*}
while for the \emph{closed system},
\begin{align*}
X_0 &=
\footnotesize
\begin{bmatrix}
0.998-0.009i & -0.005-0.000i & -0.007-0.007i\\
-0.005+0.001i & 0.999+0.009i & 0.001+0.004i\\
-0.000+0.009i & 0.015+0.011i & 1.006-0.000i\\
\end{bmatrix},\\\\
X_1 &=
\footnotesize
\begin{bmatrix}
0.003-0.002i & -0.001+0.002i & 0.001+0.001i\\
-0.001-0.003i & 0.003+0.002i & -0.000-0.000i\\
0.000-0.001i & -0.002-0.001i & 0.001-0.000i\\
\end{bmatrix},\\\\
X_2 &=
\footnotesize
\begin{bmatrix}
-0.000+0.005i & 0.002-0.001i & -0.000-0.000i\\
0.002-0.001i & -0.001-0.005i & -0.000+0.000i\\
-0.000+0.000i & 0.000+0.000i & -0.000+0.000i\\
\end{bmatrix}.
\end{align*}
\begin{equation}
    V_{3}(\epsilon; p_{1595}) = X_0 + \epsilon X_1 + \epsilon^2 X_2
\end{equation}
where for the \emph{strong noise},
\begin{align*}
X_0 &=
\footnotesize
\begin{bmatrix}
0.948-0.000i & 0.001-0.001i & 0.000+0.002i\\
-0.003-0.002i & 0.832+0.000i & 0.006-0.007i\\
0.001-0.006i & -0.006-0.007i & 0.985-0.000i\\
\end{bmatrix},\\\\
X_1 &=
\footnotesize
\begin{bmatrix}
-0.000-0.000i & -0.003-0.001i & 0.000-0.000i\\
0.010-0.004i & -0.004+0.000i & -0.001+0.001i\\
0.000+0.001i & 0.001+0.001i & 0.002+0.000i\\
\end{bmatrix},\\\\
X_2 &=
\footnotesize
\begin{bmatrix}
-0.003+0.000i & -0.002+0.002i & -0.000+0.000i\\
0.005+0.006i & -0.005+0.000i & 0.000-0.000i\\
-0.000-0.000i & -0.000-0.000i & -0.002-0.000i\\
\end{bmatrix};
\end{align*}
while for the \emph{closed system},
\begin{align*}
X_0 &=
\footnotesize
\begin{bmatrix}
1.000+0.000i & 0.002+0.003i & -0.004-0.003i\\
-0.005+0.010i & 0.994+0.000i & 0.013+0.022i\\
-0.011+0.009i & -0.013+0.022i & 0.998-0.000i\\
\end{bmatrix},\\\\
X_1 &=
\footnotesize
\begin{bmatrix}
0.010-0.000i & 0.000-0.001i & 0.001+0.001i\\
-0.001-0.002i & 0.003-0.000i & -0.002-0.004i\\
0.002-0.002i & 0.002-0.004i & 0.001+0.000i\\
\end{bmatrix},\\\\
X_2 &=
\footnotesize
\begin{bmatrix}
-0.004-0.000i & -0.002+0.000i & -0.000-0.000i\\
0.006+0.001i & -0.004-0.000i & 0.001+0.001i\\
-0.000+0.001i & -0.001+0.001i & -0.002-0.000i\\
\end{bmatrix}.
\end{align*}
\begin{equation}
    V_{4}(\epsilon; p_{1595}) = X_0 + \epsilon X_1 + \epsilon^2 X_2
\end{equation}
where for the \emph{strong noise},
\begin{align*}
X_0 &=
\footnotesize
\begin{bmatrix}
0.863-0.007i & -0.002-0.001i & 0.067-0.003i\\
-0.012+0.005i & 1.000+0.000i & -0.009+0.004i\\
0.067+0.003i & -0.005-0.002i & 0.863+0.007i\\
\end{bmatrix},\\\\
X_1 &=
\footnotesize
\begin{bmatrix}
-0.002-0.000i & 0.000+0.001i & 0.002-0.000i\\
0.002+0.001i & 0.001+0.000i & 0.001-0.001i\\
0.001+0.000i & 0.001-0.001i & -0.002+0.000i\\
\end{bmatrix},\\\\
X_2 &=
\footnotesize
\begin{bmatrix}
-0.000+0.001i & -0.000-0.000i & 0.002+0.000i\\
-0.000-0.000i & -0.002-0.000i & -0.000+0.000i\\
-0.001-0.000i & -0.000+0.000i & -0.002-0.001i\\
\end{bmatrix};
\end{align*}
while for the \emph{closed system},
\begin{align*}
X_0 &=
\footnotesize
\begin{bmatrix}
0.998-0.024i & -0.006-0.004i & -0.002-0.012i\\
-0.017+0.016i & 0.995-0.000i & -0.017+0.014i\\
0.001+0.012i & -0.006-0.006i & 1.000+0.024i\\
\end{bmatrix},\\\\
X_1 &=
\footnotesize
\begin{bmatrix}
0.004-0.006i & 0.003-0.001i & -0.005-0.003i\\
0.004-0.002i & 0.003+0.000i & 0.006+0.000i\\
0.000+0.003i & 0.001+0.001i & 0.006+0.006i\\
\end{bmatrix},\\\\
X_2 &=
\footnotesize
\begin{bmatrix}
0.001+0.002i & -0.000+0.000i & 0.001+0.001i\\
-0.001+0.000i & -0.002+0.000i & -0.001-0.000i\\
-0.002-0.001i & -0.000-0.000i & -0.001-0.002i\\
\end{bmatrix}.
\end{align*}
\begin{equation}
    V_{5}(\epsilon; p_{1595}) = X_0 + \epsilon X_1 + \epsilon^2 X_2
\end{equation}
where for the \emph{strong noise},
\begin{align*}
X_0 &=
\footnotesize
\begin{bmatrix}
0.861-0.008i & 0.006-0.008i & -0.004-0.068i\\
-0.007-0.010i & 0.996+0.000i & 0.007-0.006i\\
-0.004+0.070i & -0.009-0.007i & 0.862+0.008i\\
\end{bmatrix},\\\\
X_1 &=
\footnotesize
\begin{bmatrix}
-0.002-0.004i & -0.000+0.001i & -0.002-0.002i\\
-0.000+0.001i & 0.004+0.000i & -0.000+0.000i\\
-0.002+0.001i & 0.001+0.000i & -0.002+0.004i\\
\end{bmatrix},\\\\
X_2 &=
\footnotesize
\begin{bmatrix}
-0.001+0.004i & 0.000-0.000i & 0.002+0.001i\\
0.000-0.000i & 0.002-0.000i & 0.000-0.000i\\
0.002-0.000i & -0.000+0.000i & -0.000-0.004i\\
\end{bmatrix};
\end{align*}
while for the \emph{closed system},
\begin{align*}
X_0 &=
\footnotesize
\begin{bmatrix}
0.997-0.027i & -0.025-0.002i & -0.013+0.009i\\
0.025-0.006i & 0.996+0.000i & 0.000+0.025i\\
-0.013-0.009i & -0.004+0.025i & 0.997+0.027i\\
\end{bmatrix},\\\\
X_1 &=
\footnotesize
\begin{bmatrix}
0.013-0.005i & 0.004+0.001i & -0.003+0.004i\\
-0.004+0.003i & 0.006-0.000i & -0.000-0.005i\\
-0.003-0.008i & 0.002-0.004i & 0.011+0.005i\\
\end{bmatrix},\\\\
X_2 &=
\footnotesize
\begin{bmatrix}
-0.006+0.007i & -0.001-0.000i & 0.003-0.001i\\
0.001-0.000i & -0.002+0.000i & 0.000+0.001i\\
0.003+0.004i & -0.000+0.001i & -0.005-0.007i\\
\end{bmatrix}.
\end{align*}
\begin{equation}
    V_{6}(\epsilon; p_{1595}) = X_0 + \epsilon X_1 + \epsilon^2 X_2
\end{equation}
where for the \emph{strong noise},
\begin{align*}
X_0 &=
\footnotesize
\begin{bmatrix}
1.000+0.000i & 0.009+0.004i & -0.004-0.002i\\
-0.006+0.003i & 0.854-0.006i & 0.073-0.003i\\
0.008-0.003i & 0.073+0.003i & 0.854+0.006i\\
\end{bmatrix},\\\\
X_1 &=
\footnotesize
\begin{bmatrix}
0.004-0.000i & -0.001-0.000i & 0.000+0.001i\\
0.000-0.001i & -0.002-0.000i & 0.001-0.000i\\
-0.001+0.000i & 0.001+0.000i & -0.003+0.000i\\
\end{bmatrix},\\\\
X_2 &=
\footnotesize
\begin{bmatrix}
0.000-0.000i & 0.000+0.000i & -0.000+0.000i\\
-0.000+0.000i & -0.001-0.001i & -0.003-0.001i\\
0.000-0.000i & 0.001+0.001i & 0.001+0.001i\\
\end{bmatrix};
\end{align*}
while for the \emph{closed system},
\begin{align*}
X_0 &=
\footnotesize
\begin{bmatrix}
0.994+0.000i & 0.010-0.026i & -0.009+0.008i\\
-0.009-0.014i & 1.005-0.005i & -0.011-0.003i\\
0.010+0.020i & -0.010+0.003i & 1.006+0.005i\\
\end{bmatrix},\\\\
X_1 &=
\footnotesize
\begin{bmatrix}
0.023+0.000i & 0.002+0.007i & 0.006+0.000i\\
0.003+0.002i & 0.015-0.002i & -0.002-0.001i\\
-0.001-0.004i & -0.007+0.001i & 0.012+0.002i\\
\end{bmatrix},\\\\
X_2 &=
\footnotesize
\begin{bmatrix}
-0.003-0.000i & -0.001-0.001i & -0.002+0.000i\\
-0.001-0.001i & -0.004-0.001i & -0.003-0.000i\\
0.000+0.001i & -0.000+0.000i & -0.002+0.001i\\
\end{bmatrix}.
\end{align*}
\begin{equation}
    V_{7}(\epsilon; p_{1595}) = X_0 + \epsilon X_1 + \epsilon^2 X_2
\end{equation}
where for the \emph{strong noise},
\begin{align*}
X_0 &=
\footnotesize
\begin{bmatrix}
1.001+0.000i & 0.011+0.000i & -0.000+0.004i\\
-0.007+0.000i & 0.856-0.010i & -0.005-0.069i\\
0.000+0.009i & -0.005+0.072i & 0.858+0.010i\\
\end{bmatrix},\\\\
X_1 &=
\footnotesize
\begin{bmatrix}
-0.001-0.000i & -0.002-0.000i & -0.001-0.001i\\
0.001+0.000i & -0.005+0.002i & 0.001-0.001i\\
0.000-0.002i & 0.001+0.005i & -0.003-0.002i\\
\end{bmatrix},\\\\
X_2 &=
\footnotesize
\begin{bmatrix}
-0.003+0.000i & 0.000+0.000i & 0.000+0.000i\\
-0.000-0.000i & -0.003-0.002i & -0.001-0.001i\\
-0.000+0.000i & -0.001+0.002i & -0.002+0.002i\\
\end{bmatrix};
\end{align*}
while for the \emph{closed system},
\begin{align*}
X_0 &=
\footnotesize
\begin{bmatrix}
0.999-0.000i & 0.006-0.011i & 0.003+0.003i\\
-0.004-0.006i & 0.999-0.006i & -0.003+0.002i\\
-0.009+0.005i & -0.003+0.001i & 1.001+0.006i\\
\end{bmatrix},\\\\
X_1 &=
\footnotesize
\begin{bmatrix}
0.001-0.000i & 0.001+0.003i & 0.001-0.004i\\
0.002+0.001i & 0.000-0.000i & -0.000+0.001i\\
0.002-0.001i & -0.000-0.001i & 0.001+0.000i\\
\end{bmatrix},\\\\
X_2 &=
\footnotesize
\begin{bmatrix}
0.001-0.000i & -0.000-0.000i & -0.000+0.001i\\
-0.001-0.000i & -0.005-0.002i & -0.001-0.000i\\
-0.000+0.000i & -0.001+0.004i & -0.003+0.002i\\
\end{bmatrix}.
\end{align*}
\begin{equation}
    V_{8}(\epsilon; p_{1595}) = X_0 + \epsilon X_1 + \epsilon^2 X_2
\end{equation}
where for the \emph{strong noise},
\begin{align*}
X_0 &=
\footnotesize
\begin{bmatrix}
0.959-0.000i & 0.000+0.000i & -0.001+0.001i\\
0.000-0.000i & 0.946-0.000i & 0.006-0.002i\\
0.001+0.002i & -0.009-0.003i & 0.842+0.000i\\
\end{bmatrix},\\\\
X_1 &=
\footnotesize
\begin{bmatrix}
0.001+0.000i & 0.000+0.000i & 0.000-0.000i\\
0.000-0.000i & -0.002+0.000i & -0.000-0.000i\\
-0.000-0.000i & 0.001-0.000i & -0.006-0.000i\\
\end{bmatrix},\\\\
X_2 &=
\footnotesize
\begin{bmatrix}
-0.002+0.000i & -0.000+0.000i & 0.000+0.000i\\
-0.000-0.000i & -0.001-0.000i & 0.000-0.000i\\
-0.000+0.000i & -0.000-0.000i & 0.001-0.000i\\
\end{bmatrix};
\end{align*}
while for the \emph{closed system},
\begin{align*}
X_0 &=
\footnotesize
\begin{bmatrix}
1.001+0.000i & -0.000-0.000i & 0.001+0.006i\\
-0.000+0.000i & 1.001+0.000i & 0.006-0.003i\\
-0.001+0.010i & -0.010-0.005i & 0.998+0.000i\\
\end{bmatrix},\\\\
X_1 &=
\footnotesize
\begin{bmatrix}
0.003-0.000i & 0.004-0.001i & -0.001-0.002i\\
0.004+0.001i & 0.002-0.000i & -0.001+0.002i\\
0.001-0.003i & 0.002+0.003i & 0.004-0.000i\\
\end{bmatrix},\\\\
X_2 &=
\footnotesize
\begin{bmatrix}
-0.002-0.000i & -0.001+0.000i & 0.000+0.000i\\
-0.001-0.000i & -0.002-0.000i & 0.000-0.001i\\
-0.001+0.001i & -0.000-0.001i & -0.006-0.000i\\
\end{bmatrix}.
\end{align*}